\documentclass{llncs}
\usepackage[utf8]{inputenc}
\usepackage{booktabs}
\usepackage[colorlinks]{hyperref}

\makeatletter
\@ifundefined{lhs2tex.lhs2tex.sty.read}%
  {\@namedef{lhs2tex.lhs2tex.sty.read}{}%
   \newcommand\SkipToFmtEnd{}%
   \newcommand\EndFmtInput{}%
   \long\def\SkipToFmtEnd#1\EndFmtInput{}%
  }\SkipToFmtEnd

\newcommand\ReadOnlyOnce[1]{\@ifundefined{#1}{\@namedef{#1}{}}\SkipToFmtEnd}
\usepackage{amstext}
\usepackage{amssymb}
\usepackage{stmaryrd}
\DeclareFontFamily{OT1}{cmtex}{}
\DeclareFontShape{OT1}{cmtex}{m}{n}
  {<5><6><7><8>cmtex8
   <9>cmtex9
   <10><10.95><12><14.4><17.28><20.74><24.88>cmtex10}{}
\DeclareFontShape{OT1}{cmtex}{m}{it}
  {<-> ssub * cmtt/m/it}{}

\DeclareFontShape{OT1}{cmtt}{bx}{n}
  {<5><6><7><8>cmtt8
   <9>cmbtt9
   <10><10.95><12><14.4><17.28><20.74><24.88>cmbtt10}{}
\DeclareFontShape{OT1}{cmtex}{bx}{n}
  {<-> ssub * cmtt/bx/n}{}

\newcommand{\Conid}[1]{\mathit{#1}}
\newcommand{\Varid}[1]{\mathit{#1}}
\newcommand{\anonymous}{\kern0.06em \vbox{\hrule\@width.5em}}
\newcommand{\plus}{\mathbin{+\!\!\!+}}

\newcommand{\sequ}{\mathbin{>\!\!\!>}}
\renewcommand{\leq}{\leqslant}
\renewcommand{\geq}{\geqslant}
\usepackage{polytable}

\@ifundefined{mathindent}%
  {\newdimen\mathindent\mathindent\leftmargini}%
  {}%

\def\resethooks{%
  \global\let\SaveRestoreHook\empty
  \global\let\ColumnHook\empty}
\newcommand*{\savecolumns}[1][default]%
  {\g@addto@macro\SaveRestoreHook{\savecolumns[#1]}}
\newcommand*{\restorecolumns}[1][default]%
  {\g@addto@macro\SaveRestoreHook{\restorecolumns[#1]}}
\newcommand*{\aligncolumn}[2]%
  {\g@addto@macro\ColumnHook{\column{#1}{#2}}}

\resethooks

\newcommand{\onelinecommentchars}{\quad-{}- }
\newcommand{\commentbeginchars}{\enskip\{-}
\newcommand{\commentendchars}{-\}\enskip}

\newcommand{\visiblecomments}{%
  \let\onelinecomment=\onelinecommentchars
  \let\commentbegin=\commentbeginchars
  \let\commentend=\commentendchars}

\newcommand{\invisiblecomments}{%
  \let\onelinecomment=\empty
  \let\commentbegin=\empty
  \let\commentend=\empty}

\visiblecomments

\newlength{\blanklineskip}
\newcommand{\hsindent}[1]{\quad}
\let\hspre\empty
\let\hspost\empty

\EndFmtInput
\makeatother
\ReadOnlyOnce{polycode.fmt}%
\makeatletter

\newcommand{\hsnewpar}[1]%
  {{\parskip=0pt\parindent=0pt\par\vskip #1\noindent}}

\newcommand{\hscodestyle}{}

\newcommand{\sethscode}[1]%
  {\expandafter\let\expandafter\hscode\csname #1\endcsname
   \expandafter\let\expandafter\endhscode\csname end#1\endcsname}

  {\par\noindent
   \advance\leftskip\mathindent
   \hscodestyle
   \let\\=\@normalcr
   \let\hspre\(\let\hspost\)%
   \pboxed}%
  {\endpboxed\)%
   \par\noindent
   \ignorespacesafterend}

  {\hsnewpar\abovedisplayskip
   \advance\leftskip\mathindent
   \hscodestyle
   \let\hspre\(\let\hspost\)%
   \pboxed}%
  {\endpboxed%
   \hsnewpar\belowdisplayskip
   \ignorespacesafterend}

  {\hsnewpar\abovedisplayskip
   \advance\leftskip\mathindent
   \hscodestyle
   \let\\=\@normalcr
   \(\pboxed}%
  {\endpboxed\)%
   \hsnewpar\belowdisplayskip
   \ignorespacesafterend}

\newcommand{\plainhs}{\sethscode{plainhscode}}

\plainhs

  {\hsnewpar\abovedisplayskip
   \advance\leftskip\mathindent
   \hscodestyle
   \let\\=\@normalcr
   \(\parray}%
  {\endparray\)%
   \hsnewpar\belowdisplayskip
   \ignorespacesafterend}

  {\parray}{\endparray}

  {\(\parray}{\endparray\)}

\def\codeframewidth{\arrayrulewidth}
\RequirePackage{calc}

  {\parskip=\abovedisplayskip\par\noindent
   \hscodestyle
   \arrayrulewidth=\codeframewidth
   \tabular{@{}|p{\linewidth-2\arraycolsep-2\arrayrulewidth-2pt}|@{}}%
   \hline\framedhslinecorrect\\{-1.5ex}%
   \let\endoflinesave=\\
   \let\\=\@normalcr
   \(\pboxed}%
  {\endpboxed\)%
   \framedhslinecorrect\endoflinesave{.5ex}\hline
   \endtabular
   \parskip=\belowdisplayskip\par\noindent
   \ignorespacesafterend}

\newcommand{\framedhslinecorrect}[2]%
  {#1[#2]}

  {\(\def\column##1##2{}%
   \let\>\undefined\let\<\undefined\let\\\undefined
   \newcommand\>[1][]{}\newcommand\<[1][]{}\newcommand\\[1][]{}%
   \def\fromto##1##2##3{##3}%
   }{\) }%

  {\let\orighscode=\hscode
   \let\origendhscode=\endhscode
   \def\endhscode{\def\hscode{\endgroup\def\@currenvir{hscode}\\}\begingroup}
   \orighscode\def\hscode{\endgroup\def\@currenvir{hscode}}}%
  {\origendhscode
   \global\let\hscode=\orighscode
   \global\let\endhscode=\origendhscode}%

\makeatother
\EndFmtInput

\renewcommand{\hscodestyle}{\small}

\usepackage{xcolor}
\colorlet{rev}{black}
\newcommand\REV[1]{{\color{rev}#1}}
\newenvironment{REVENV}{\color{rev}}{}

\title{Practical Idiomatic Considerations for Checkable~Meta-Logic in Experimental Functional~Programming}
\author{Baltasar Trancón y Widemann \and Markus Lepper}
\institute{semantics GmbH, Berlin}

\begin{document}
\maketitle

\begin{abstract}
  Implementing a complex concept as an executable model in a strongly typed,
  purely functional language hits a sweet spot between mere simulation and
  formal specification.  For research and education it is often desirable to
  enrich the algorithmic code with meta-logical annotations, variously embodied
  as assertions, theorems or test cases.  Checking frameworks use the inherent
  logical power of the functional paradigm to approximate theorem proving by
  heuristic testing.  Here we propose several novel idioms to enhance the
  practical expressivity of checking, namely meta-language marking, nominal
  axiomatics, and constructive existentials.  All of these are formulated in
  literate Haskell'98 with some common language extensions.  Their use and
  impact are illustrated by application to a realistic modeling problem.
\end{abstract}


\section{Introduction}

This paper discusses general programming methodology in terms of a particular
implementation in Haskell.  Thus it is provided as a literate
Haskell\,\cite{literate} program.\footnote{\REV{A full and self-contained source
archive for practical evaluation is publicly available at
\url{http://bandm.eu/download/purecheck/}.}}

\subsection{Proving and Checking}

Purely functional programming has arguably a friendlier relationship to
meta-logic, the discipline of formal reasoning about program properties, than
conventional state-based paradigms \cite{Backus1978}.  This has been exploited
in a number of ways that differ greatly in their pragmatic context.

At one end of the spectrum, strongly normalizing languages and the
types-as-propositions approach, ultimately based on the
Brouwer--Heyting--Kolmogorov interpretation of constructive logic, have led to
the unification of algorithmic programming and constructive \emph{theorem
proving}.  The practice has evolved from basic models such as the Calculus of
Constructions \cite{coc} to full-blown languages and interactive programming
environments such as Agda \cite{agda}.  The basic approach is that a program is
statically validated with respect to a type signature that encodes the
desired meta-logical property, if and only if it truly possesses that property.
This approach does evidently not scale to Turing-complete languages.\,\footnote{Consider
the ``constructive'' logical reading of the type of a generic recursion
operator, \ensuremath{(\alpha\to \alpha)\to \alpha}; it says literally that \REV{\emph{begging the
question}, \ensuremath{\alpha\to \alpha}, is a valid proof method for any proposition \ensuremath{\alpha}}.}  Thus, for meta-logic over complete
programming languages, it is not sufficient to demonstrate \emph{inhabitation}
of a type to obtain a proof, but it must also stand the test of successful
\emph{evaluation}.

At the other end of the spectrum, freedom from side effects allows for liberally
sprinkling program code with \emph{online assertions}, that is, computations
whose values the program depends on not for its outcome, but for ensuring its
correct operation.  Some care must be exercised when timing and strictness
details matter \cite{hask-assert}, but otherwise the technique is just as
powerful as for conventional programming paradigms \cite{assert}, minus the need
for a pure assertion sublanguage.

The middle ground is covered by \emph{offline checking}\footnote{\REV{Thus named here for clear
contrast with the alternatives, but largely synonymous with \emph{property-based
testing}\,\cite{prop}.}}, that is, evaluation of meta-logical properties
\REV{as a separate mode of program} execution.  Offline checking is of course less
rigorous than theorem proving, and may involve incomplete and heuristic
reasoning procedures.  On the other hand, it is more abstract and static than
online assertions; thus cases that are not reached during online evaluation can
be covered, and the checking effort can be shifted to convenient points in the
software lifecycle.  Offline checking fills the same role as conventional
\emph{unit testing} procedures, although the focus is a bit different:
\REV{checking purely functional programs} is \REV{commonly} both simpler in control, due to the lack of state of the
unit under test that needs to be set up and observed, and more complex in data,
due to the pervasiveness of higher-order functions.

There are various popular offline checking frameworks for functional programming
languages, \REV{such as QuickCheck, (Lazy) SmallCheck, SmartCheck, ScalaCheck or PropEr,} and we
assume the reader is familiar with their general design and operation, for
instance with the seminal QuickCheck \cite{quickcheck} for Haskell.

\subsection{Executable Modeling}

The field of executable modeling, that is, the construction of experimental
programs that embody theoretical concepts of systems and processes, and imbue
them with practically observable behavior, poses specific challenges.  In
particular, some mechanism is needed to \emph{validate} the implementation, that
is, establish trust in its faithful representation of the concepts under study.
Since executable models are designed to \REV{exceed behavioral a-priori intuition} (otherwise their content were
trivial)\cite{virtual}, it is intrinsically hard to differentiate bugs from features.

In a naive idealistic sense, model programs should be derived and proved
rigorously.  However, that presupposes a complete, \REV{computable and}
operationalized theory.  For the two scenarios where a theory exists but is not
fully operationalized, and where models are used \emph{inductively} as
approximations to a future theory, we consider less rigorous approaches, and
offline checking in particular, the more viable validation procedure.  It may
even be educational to both check and run models that behave evidently wrong.

\subsection{The PureCheck Framework}

The checking idioms to be proposed in the following have been developed in the
context of an experimental checking framework, PureCheck, implemented as a plain
Haskell library.  The design of PureCheck largely follows the paradigm of
popular frameworks such as QuickCheck\cite{quickcheck} or
SmallCheck\cite{smallcheckx}, with some notable deviations.

Like other frameworks, PureCheck leverages the internal logical language of
functional programming, and type-directed generation of test data for universal
propositions.  PureCheck prioritizes purity and non-strictness; text execution
is rigidly non-monadic, and thus equally suitable for both offline checks and
online assertions.  Unlike QuickCheck, test data are generated by deterministic
rather than randomized combinatorial procedures.  \REV{Unlike SmallCheck, sample
sizes can be bounded precisely, without risk of combinatorial explosion.}  Test
data sets are pessimistically assumed to be possibly insufficient, and thus the
direction of logical approximation is significant; evaluation may yield false
positives, resulting from undiscovered counterexamples, but never false
negatives.

\begin{REVENV}
The contribution of the present paper is a collection of three novel and
experimental idioms for offline checking.  The definitions and an example
application are given in the following two main sections, respectively.  These
features have been implemented in Haskell for PureCheck, but are theoretically
compatible with other frameworks and host languages.
\end{REVENV}

\subsubsection{PureCheck Basics}

At the heart of the framework is an encapsulation for heuristic checks.
\begin{hscode}\SaveRestoreHook
\column{B}{@{}>{\hspre}l<{\hspost}@{}}%
\column{3}{@{}>{\hspre}l<{\hspost}@{}}%
\column{E}{@{}>{\hspre}l<{\hspost}@{}}%
\>[3]{}\mathbf{newtype}\;\Conid{Check}\mathrel{=}\Conid{Check}\;\{\mskip1.5mu \Varid{perform}\mathbin{::}\Conid{Int}\to \Conid{Bool}\mskip1.5mu\}{}\<[E]%
\ColumnHook
\end{hscode}\resethooks
The heuristic is parameterized with an \ensuremath{\Conid{Int}} called the \emph{confidence}
parameter.  Because of monotonicity, higher values may require more
computational effort, but can only improve the test accuracy by eliminating more
false positives.

The propositions that can be encapsulated in this way come in various shapes;
thus we define a type class with an ad-hoc polymorphic encapsulation operation.
\begin{hscode}\SaveRestoreHook
\column{B}{@{}>{\hspre}l<{\hspost}@{}}%
\column{3}{@{}>{\hspre}l<{\hspost}@{}}%
\column{E}{@{}>{\hspre}l<{\hspost}@{}}%
\>[3]{}\mathbf{class}\;\Conid{Checkable}\;\alpha\;\mathbf{where}\;\Varid{check}\mathbin{::}\Conid{Meta}\;\alpha\to \Conid{Check}{}\<[E]%
\ColumnHook
\end{hscode}\resethooks
The wrapper \ensuremath{\Conid{Meta}} should be ignored for now; it shall be discussed in due
detail in the following section.  The base case is a propositional constant.
\begin{hscode}\SaveRestoreHook
\column{B}{@{}>{\hspre}l<{\hspost}@{}}%
\column{3}{@{}>{\hspre}l<{\hspost}@{}}%
\column{E}{@{}>{\hspre}l<{\hspost}@{}}%
\>[3]{}\mathbf{instance}\;\Conid{Checkable}\;\Conid{Bool}\;\mathbf{where}\;\Varid{check}\;(\Conid{Meta}\;\Varid{b})\mathrel{=}\Conid{Check}\;(\Varid{const}\;\Varid{b}){}\<[E]%
\ColumnHook
\end{hscode}\resethooks
Checks bear the obvious conjunctive monoid structure.  Since the aggregate
confidence in the truth of a conjunction can be no higher than the individual
confidence in any of its clauses, the parameter is copied clause-wise.
\begin{hscode}\SaveRestoreHook
\column{B}{@{}>{\hspre}l<{\hspost}@{}}%
\column{3}{@{}>{\hspre}l<{\hspost}@{}}%
\column{5}{@{}>{\hspre}l<{\hspost}@{}}%
\column{34}{@{}>{\hspre}c<{\hspost}@{}}%
\column{34E}{@{}l@{}}%
\column{37}{@{}>{\hspre}l<{\hspost}@{}}%
\column{E}{@{}>{\hspre}l<{\hspost}@{}}%
\>[3]{}\mathbf{instance}\;\Conid{Monoid}\;\Conid{Check}\;\mathbf{where}{}\<[E]%
\\
\>[3]{}\hsindent{2}{}\<[5]%
\>[5]{}\Varid{mempty}{}\<[34]%
\>[34]{}\mathrel{=}{}\<[34E]%
\>[37]{}\Conid{Check}\;(\lambda \Varid{n}\to \Conid{True}){}\<[E]%
\\
\>[3]{}\hsindent{2}{}\<[5]%
\>[5]{}\Varid{mappend}\;(\Conid{Check}\;\Varid{c})\;(\Conid{Check}\;\Varid{d}){}\<[34]%
\>[34]{}\mathrel{=}{}\<[34E]%
\>[37]{}\Conid{Check}\;(\lambda \Varid{n}\to \Varid{c}\;\Varid{n}\mathrel{\wedge}\Varid{d}\;\Varid{n}){}\<[E]%
\\[\blanklineskip]%
\>[3]{}\mathbf{instance}\;\Conid{Checkable}\;()\;\mathbf{where}\;\Varid{check}\;(\Conid{Meta}\;())\mathrel{=}\Varid{mempty}{}\<[E]%
\\[\blanklineskip]%
\>[3]{}\mathbf{instance}\;(\Conid{Checkable}\;\alpha,\Conid{Checkable}\;\beta)\Rightarrow \Conid{Checkable}\;(\alpha,\beta)\;\mathbf{where}{}\<[E]%
\\
\>[3]{}\hsindent{2}{}\<[5]%
\>[5]{}\Varid{check}\;(\Conid{Meta}\;(\Varid{p},\Varid{q}))\mathrel{=}\Varid{check}\;(\Conid{Meta}\;\Varid{p})\mathbin{`\Varid{mappend}`}\Varid{check}\;(\Conid{Meta}\;\Varid{q}){}\<[E]%
\\[\blanklineskip]%
\>[3]{}\mathbf{instance}\;(\Conid{Checkable}\;\alpha)\Rightarrow \Conid{Checkable}\;[\mskip1.5mu \alpha\mskip1.5mu]\;\mathbf{where}{}\<[E]%
\\
\>[3]{}\hsindent{2}{}\<[5]%
\>[5]{}\Varid{check}\;(\Conid{Meta}\;\Varid{ps})\mathrel{=}\Varid{mconcat}\;(\Varid{map}\;(\Varid{check}\mathbin{\circ}\Conid{Meta})\;\Varid{ps}){}\<[E]%
\ColumnHook
\end{hscode}\resethooks

For quantified universals, a generator for representative samples of the
argument space is required.  The confidence parameter is taken as the
recommended maximum sample size \REV{(unlike SmallCheck, where the parameter is
a \emph{depth} to be exhausted, such that sample size may be only exponentially related)}.
Unlike in the conjunctive case, nested universal quantifiers are not simply
dealt with recursively.  Instead, it is recommended to use uncurried forms
quantified over tuples to ensure proper weight-balancing between argument
samples.
\begin{hscode}\SaveRestoreHook
\column{B}{@{}>{\hspre}l<{\hspost}@{}}%
\column{3}{@{}>{\hspre}l<{\hspost}@{}}%
\column{5}{@{}>{\hspre}l<{\hspost}@{}}%
\column{E}{@{}>{\hspre}l<{\hspost}@{}}%
\>[3]{}\Varid{checkWith}\mathbin{::}\Conid{Generator}\;\alpha\to \Conid{Meta}\;(\alpha\to \Conid{Bool})\to \Conid{Check}{}\<[E]%
\\
\>[3]{}\Varid{checkWith}\;\Varid{g}\;(\Conid{Meta}\;\Varid{p})\mathrel{=}\Conid{Check}\;(\lambda \Varid{n}\to \Varid{all}\;\Varid{p}\;(\Varid{generate}\;\Varid{g}\;\Varid{n})){}\<[E]%
\\[\blanklineskip]%
\>[3]{}\mathbf{instance}\;(\Conid{Some}\;\alpha)\Rightarrow \Conid{Checkable}\;(\alpha\to \Conid{Bool})\;\mathbf{where}{}\<[E]%
\\
\>[3]{}\hsindent{2}{}\<[5]%
\>[5]{}\Varid{check}\mathrel{=}\Varid{checkWith}\;\Varid{some}{}\<[E]%
\ColumnHook
\end{hscode}\resethooks

Test data generators are wrapped pure functions, and thus deterministic in the
size parameter \ensuremath{\Varid{n}}.  Useful generators return at most (preferably approximately)
\ensuremath{\Varid{n}} elements (preferably distinct and with commensurate internal variety).
\begin{hscode}\SaveRestoreHook
\column{B}{@{}>{\hspre}l<{\hspost}@{}}%
\column{3}{@{}>{\hspre}l<{\hspost}@{}}%
\column{E}{@{}>{\hspre}l<{\hspost}@{}}%
\>[3]{}\mathbf{newtype}\;\Conid{Generator}\;\alpha\mathrel{=}\Conid{Generator}\;\{\mskip1.5mu \Varid{generate}\mathbin{::}\Conid{Int}\to [\mskip1.5mu \alpha\mskip1.5mu]\mskip1.5mu\}{}\<[E]%
\ColumnHook
\end{hscode}\resethooks
A type class provides default generators for its instance types.
\begin{hscode}\SaveRestoreHook
\column{B}{@{}>{\hspre}l<{\hspost}@{}}%
\column{3}{@{}>{\hspre}l<{\hspost}@{}}%
\column{E}{@{}>{\hspre}l<{\hspost}@{}}%
\>[3]{}\mathbf{class}\;\Conid{Some}\;\alpha\;\mathbf{where}\;\Varid{some}\mathbin{::}\Conid{Generator}\;\alpha{}\<[E]%
\ColumnHook
\end{hscode}\resethooks
\begin{REVENV}
Generators for simple types are straightforward, for instance:
\begin{hscode}\SaveRestoreHook
\column{B}{@{}>{\hspre}l<{\hspost}@{}}%
\column{3}{@{}>{\hspre}l<{\hspost}@{}}%
\column{5}{@{}>{\hspre}l<{\hspost}@{}}%
\column{E}{@{}>{\hspre}l<{\hspost}@{}}%
\>[3]{}\mathbf{instance}\;\Conid{Some}\;\Conid{Bool}\;\mathbf{where}{}\<[E]%
\\
\>[3]{}\hsindent{2}{}\<[5]%
\>[5]{}\Varid{some}\mathrel{=}\Conid{Generator}\mathbin{\$}\Varid{flip}\;\Varid{take}\;[\mskip1.5mu \Conid{False},\Conid{True}\mskip1.5mu]{}\<[E]%
\ColumnHook
\end{hscode}\resethooks
Generator combinators for complex types need to consider the issues of weight
balancing between dimensions and of infinite enumerations;  the details are out
of scope here.\footnote{\REV{Implementations can be found in the full source.}}
\end{REVENV}

\section{Proposed Idioms}

\subsection{Meta-Language Marking}

The principle of types as propositions in a functional programming language is a
two-sided coin.  On the upside, the internal logical language is automatically
consistent with the language semantics, and quite expressive.  On the downside,
the expressive power of advanced abstractions such as higher-order functions and
polymorphism is a bit too much for the logical needs of the average user.
Unrestrained use can make the meta-logical aspects of the codebase
overwhelmingly hard to both write and read.\,\footnote{The reader is invited to
contemplate for example the variety of possible higher-order logical meanings of
the following specialization of a well-known Haskell Prelude function:
\ensuremath{\Varid{foldl}\mathbin{::}(\Conid{Foldable}\;\tau)\Rightarrow (\Conid{Bool}\to \alpha\to \Conid{Bool})\to \Conid{Bool}\to \tau\;\alpha\to \Conid{Bool}}}

We propose that, for both education and engineering, it is a wise move to
delimit the parts of the codebase that are intended as meta-logical vocabulary
explicitly.  To this end, we introduce a generic wrapper type.
\begin{hscode}\SaveRestoreHook
\column{B}{@{}>{\hspre}l<{\hspost}@{}}%
\column{3}{@{}>{\hspre}l<{\hspost}@{}}%
\column{E}{@{}>{\hspre}l<{\hspost}@{}}%
\>[3]{}\mathbf{data}\;\Conid{Meta}\;\alpha\mathrel{=}\Conid{Meta}\;\{\mskip1.5mu \Varid{reflect}\mathbin{::}\alpha\mskip1.5mu\}{}\<[E]%
\ColumnHook
\end{hscode}\resethooks
Then the codebase is manifestly stratified into four layers:
\begin{description}\small

\item[Operational] definitions do \emph{not} use the \ensuremath{\Conid{Meta}} type/value
  constructor.

\item[Assertive] definitions use the \ensuremath{\Conid{Meta}} constructor in \emph{root} position.

\item[Tactical] definitions use the \ensuremath{\Conid{Meta}} constructor in \emph{non-root}
  position.

\item[Transcendent] definitions are polymorphic over a type (constructor)
  variable that admits some \ensuremath{\Conid{Meta}\;\alpha} (or \ensuremath{\Conid{Meta}} itself, respectively) as an
  \emph{instance}.

\end{description}

The effect of this stratified marking discipline is that, contrarily to the
pathological \ensuremath{\Varid{foldl}} example presented above, the intented reading of type
signatures becomes clear.  For instance:
\begin{itemize}\small

\item \ensuremath{\Conid{Meta}\;\Conid{Bool}} is the type of atomic assertive meta-expressions that are
  expected to evaluate straightforwardly to \ensuremath{\Conid{True}}; definitions of this type
  incur a \emph{singleton} static checking obligation, that is a \emph{test
  case}.

\item \ensuremath{\Conid{Meta}\;\Conid{Int}} is a type of meta-expressions without a truth-value, let alone
  an expected one; definitions of this type incur no static checking obligation.

\item \ensuremath{\Conid{Meta}\;(\Conid{A}\to \Conid{Bool})}\footnote{Note the discourse-level meta-variable \ensuremath{\Conid{A}} for
  a monomorphic Haskell type, instead of an object-level type variable \ensuremath{\alpha}.}
  is the type of quantified assertive meta-expressions that are expected to
  evaluate to \ensuremath{\Conid{True}} for all parameter values of type \ensuremath{\Conid{A}}; definitions of this
  type incur a static checking obligation for \emph{some} values (preferably a
  representative set).

\item \ensuremath{\Conid{A}\to \Conid{Meta}\;\Conid{Bool}} is the type of tactics that can construct such assertions
  from parameter values of type \ensuremath{\Conid{A}}; definitions of this type incur \emph{no}
  static checking obligation, but may implement an aspect of a \emph{test
  strategy}.

\item \ensuremath{\Conid{Meta}\;\Conid{A}\to \Conid{Meta}\;\Conid{B}} is the type of tactics that can transform an assertion
  of type \ensuremath{\Conid{A}} to an assertion of type \ensuremath{\Conid{B}}; definitions of this type incur no
  checking obligation, but may implement an aspect of a test strategy.

\item \ensuremath{\Conid{Meta}\;\Conid{A}\to \Conid{Bool}} is the type of tactics that can evaluate a meta-property
  of an assertion of type \ensuremath{\Conid{A}}; definitions of this type incur no checking
  obligation, but may implement an aspect of a test strategy.

\item \ensuremath{\Conid{Meta}\;(\alpha\to \Conid{Bool})\to \Conid{Meta}\;([\mskip1.5mu \alpha\mskip1.5mu]\to \Conid{Bool})} is the type of
  parametrically polymorphic predicate transformers that lifts an assertion
  meta-expression quantified over an arbitrary element type \ensuremath{\alpha} to one
  quantified over the corresponding list type \ensuremath{[\mskip1.5mu \alpha\mskip1.5mu]}.

\end{itemize}

Note that transport of operational subexpressions into the meta-logical layer is
the simple matter of a \ensuremath{\Conid{Meta}} data constructor.  By contrast, the reverse
transport using the projection \ensuremath{\Varid{reflect}} is discouraged except for certain
idiomatic cases.

Evidently level marking makes no contribution to algorithmic computations.  That
it is pragmatically valuable documentation nevertheless is demonstrated by the
explicit meta-logical universal quantifier:
\begin{hscode}\SaveRestoreHook
\column{B}{@{}>{\hspre}l<{\hspost}@{}}%
\column{3}{@{}>{\hspre}l<{\hspost}@{}}%
\column{E}{@{}>{\hspre}l<{\hspost}@{}}%
\>[3]{}\Varid{foreach}\mathbin{::}(\alpha\to \Conid{Meta}\;\beta)\to \Conid{Meta}\;(\alpha\to \beta){}\<[E]%
\\
\>[3]{}\Varid{foreach}\;\Varid{f}\mathrel{=}\Conid{Meta}\;(\lambda \Varid{x}\to \Varid{reflect}\;(\Varid{f}\;\Varid{x})){}\<[E]%
\ColumnHook
\end{hscode}\resethooks
If \ensuremath{\Varid{f}} is a predicate that is used pointwise to form meta-expressions, then
\ensuremath{\Varid{foreach}\;\Varid{f}} is a singular meta-expression that quantifies over all points.  For
example,
\begin{hscode}\SaveRestoreHook
\column{B}{@{}>{\hspre}l<{\hspost}@{}}%
\column{3}{@{}>{\hspre}l<{\hspost}@{}}%
\column{E}{@{}>{\hspre}l<{\hspost}@{}}%
\>[3]{}\Conid{Meta}\mathbin{\circ}\Varid{even}\mathbin{::}\Conid{Int}\to \Conid{Meta}\;\Conid{Bool}{}\<[E]%
\ColumnHook
\end{hscode}\resethooks
is clearly a predicate intended to be used pointwise since the alternative
reading, ``all integers are even'', is blatantly false.  By contrast,
\begin{hscode}\SaveRestoreHook
\column{B}{@{}>{\hspre}l<{\hspost}@{}}%
\column{3}{@{}>{\hspre}l<{\hspost}@{}}%
\column{E}{@{}>{\hspre}l<{\hspost}@{}}%
\>[3]{}\Varid{foreach}\;(\Conid{Meta}\mathbin{\circ}\Varid{even}\mathbin{\circ}(\mathbin{*}\mathrm{2}))\mathbin{::}\Conid{Meta}\;(\Conid{Int}\to \Conid{Bool}){}\<[E]%
\ColumnHook
\end{hscode}\resethooks
is a (true) universal assertion quantified over all (non-\ensuremath{\bot }) values of
type \ensuremath{\Conid{Int}}.

\begin{REVENV} As a more relevant example, consider a preorder of meta-logical
interest, say a semantic approximation relation, on some data type \ensuremath{\Conid{A}}.
\begin{hscode}\SaveRestoreHook
\column{B}{@{}>{\hspre}l<{\hspost}@{}}%
\column{3}{@{}>{\hspre}l<{\hspost}@{}}%
\column{E}{@{}>{\hspre}l<{\hspost}@{}}%
\>[3]{}(\mathrel{\sqsubseteq})\mathbin{::}\Conid{A}\to \Conid{A}\to \Conid{Meta}\;\Conid{Bool}{}\<[E]%
\ColumnHook
\end{hscode}\resethooks
This is directly usable as a binary predicate that characterizes the
relationship of two particular elements.  By converting one quantifier, we
obtain a unary predicate that characterizes a particular element as globally
minimal:
\begin{hscode}\SaveRestoreHook
\column{B}{@{}>{\hspre}l<{\hspost}@{}}%
\column{3}{@{}>{\hspre}l<{\hspost}@{}}%
\column{E}{@{}>{\hspre}l<{\hspost}@{}}%
\>[3]{}\Varid{minimal}\mathbin{::}\Conid{A}\to \Conid{Meta}\;(\Conid{A}\to \Conid{Bool}){}\<[E]%
\\
\>[3]{}\Varid{minimal}\;\Varid{x}\mathrel{=}\Varid{foreach}\;(\Varid{x}\mathrel{\sqsubseteq}){}\<[E]%
\ColumnHook
\end{hscode}\resethooks
By converting the other quantifier also, we obtain a nullary predicate that
characterizes the preorder as trivial:
\begin{hscode}\SaveRestoreHook
\column{B}{@{}>{\hspre}l<{\hspost}@{}}%
\column{3}{@{}>{\hspre}l<{\hspost}@{}}%
\column{E}{@{}>{\hspre}l<{\hspost}@{}}%
\>[3]{}\Varid{trivial}\mathbin{::}\Conid{Meta}\;(\Conid{A}\to \Conid{A}\to \Conid{Bool}){}\<[E]%
\\
\>[3]{}\Varid{trivial}\mathrel{=}\Varid{foreach}\;\Varid{minimal}{}\<[E]%
\ColumnHook
\end{hscode}\resethooks
The final conversion to the recommended uncurried type \ensuremath{\Conid{Meta}\;((\Conid{A},\Conid{A})\to \Conid{Bool})}
can be performed explicitly (left as an exercise to the reader), or implicitly
by a suitable instance of \ensuremath{\Conid{Checkable}}.

This style ensures that higher-order functions and meta-logical reading are
orthogonal means of expressivity.
\end{REVENV}

All checking ultimately involves the evaluation of an expression of type \ensuremath{\Conid{Meta}\;\Conid{Bool}}.  The denotational semantics of this \REV{Haskell} type has four meaningful values,
namely:

\medskip
\begin{center}
  \def\arraystretch{1.1}\small
  \tabcolsep=1em
  \begin{tabular}{lll}
    \toprule
    \bfseries Value & {\bfseries Verdict:} The checked property & \bfseries Issue Type
    \\ \midrule
    \ensuremath{\Conid{Meta}\;\Conid{True}} & \dots holds & \enspace ---
    \\
    \ensuremath{\Conid{Meta}\;\Conid{False}} & \dots does not hold & logical falsehood
    \\
    \ensuremath{\Conid{Meta}\;\bot } & \dots cannot be decided & logical error
    \\
    \ensuremath{\bot } & \dots cannot be stated & tactical error
    \\ \bottomrule
  \end{tabular}
\end{center}

Semantic \ensuremath{\bot } values occuring intermediately, such as in tactical
computations or test data generation, are not constrained by our framework.  To
the contrary, non-strictness can be exploited in useful ways to manipulate
complex meta-logical constructs.  For instance, consider a form of bounded
quantification, where an explicit sample generator is provided:
\begin{hscode}\SaveRestoreHook
\column{B}{@{}>{\hspre}l<{\hspost}@{}}%
\column{3}{@{}>{\hspre}l<{\hspost}@{}}%
\column{5}{@{}>{\hspre}l<{\hspost}@{}}%
\column{E}{@{}>{\hspre}l<{\hspost}@{}}%
\>[3]{}\mathbf{data}\;\Conid{For}\;\alpha\;\beta\mathrel{=}\Conid{For}\;\{\mskip1.5mu \Varid{bound}\mathbin{::}\Conid{Generator}\;\alpha,\Varid{body}\mathbin{::}\alpha\to \beta\mskip1.5mu\}{}\<[E]%
\\[\blanklineskip]%
\>[3]{}\mathbf{instance}\;\Conid{Checkable}\;(\Conid{For}\;\alpha\;\Conid{Bool})\;\mathbf{where}{}\<[E]%
\\
\>[3]{}\hsindent{2}{}\<[5]%
\>[5]{}\Varid{check}\;(\Conid{Meta}\;(\Conid{For}\;\Varid{g}\;\Varid{p}))\mathrel{=}\Varid{checkWith}\;\Varid{g}\;(\Conid{Meta}\;\Varid{p}){}\<[E]%
\ColumnHook
\end{hscode}\resethooks
Nested bounded quantifications of the form \ensuremath{\Conid{For}\;\Varid{g}\;(\lambda \Varid{x}\to \Conid{For}\;\Varid{h}\;(\lambda \Varid{y}\to \Varid{p}))}
cannot be merged or transposed straightforwardly, because a lambda abstraction
intervenes.  However, semantics can be exploited if \ensuremath{\Varid{h}} is independent of, and
thus non-strict in \ensuremath{\Varid{x}}.
\begin{hscode}\SaveRestoreHook
\column{B}{@{}>{\hspre}l<{\hspost}@{}}%
\column{3}{@{}>{\hspre}l<{\hspost}@{}}%
\column{21}{@{}>{\hspre}c<{\hspost}@{}}%
\column{21E}{@{}l@{}}%
\column{24}{@{}>{\hspre}l<{\hspost}@{}}%
\column{29}{@{}>{\hspre}l<{\hspost}@{}}%
\column{32}{@{}>{\hspre}c<{\hspost}@{}}%
\column{32E}{@{}l@{}}%
\column{35}{@{}>{\hspre}l<{\hspost}@{}}%
\column{E}{@{}>{\hspre}l<{\hspost}@{}}%
\>[3]{}\Varid{qmerge}\mathbin{::}\Conid{For}\;\alpha\;(\Conid{For}\;\beta\;\gamma)\to \Conid{For}\;(\alpha,\beta)\;\gamma{}\<[E]%
\\
\>[3]{}\Varid{qmerge}\;(\Conid{For}\;\Varid{g}\;\Varid{k}){}\<[21]%
\>[21]{}\mathrel{=}{}\<[21E]%
\>[24]{}\mathbf{let}\;{}\<[29]%
\>[29]{}\Varid{h}{}\<[32]%
\>[32]{}\mathrel{=}{}\<[32E]%
\>[35]{}\Varid{bound}\;(\Varid{k}\;\bot ){}\<[E]%
\\
\>[24]{}\mathbf{in}\;{}\<[29]%
\>[29]{}\Conid{For}\;(\Varid{gpair}\;\Varid{g}\;\Varid{h})\;(\lambda (\Varid{x},\Varid{y})\to \Varid{body}\;(\Varid{k}\;\Varid{x})\;\Varid{y}){}\<[E]%
\ColumnHook
\end{hscode}\resethooks
Here \ensuremath{\Varid{gpair}} forms a Cartesian sample product for marginal generators \ensuremath{\Varid{g}} and
\ensuremath{\Varid{h}}.\footnote{\REV{The implementation of \ensuremath{\Varid{gpair}} is explained in detail in the full source.}}

\subsection{Nominal Axiomatics}

In a types-as-propositions approach to meta-logic of functional programs, a
property of interest is encoded as a dependent type, and holds if the type can
be demonstrated to be inhabited in a constructive semantics.

By contrast, checking approaches are \emph{empirical}: Properties of interest
are tested by computable functions, and thus collapse to the result type \ensuremath{\Conid{Bool}},
of which only the value \ensuremath{\Conid{True}} is accepted.  A seemingly trivial, but
practically significant consequence is that type signatures are not helpful to
prevent accidental confusion of structurally similar properties.

This issue is compounded, quite paradoxically, by abstraction mechanisms.  Often
a proposition can be stated in concise generic form by abstraction from values,
types or type class instances.  The actual checking then operates on a
particular concretization (by application in the former and type inference in
the latter two cases, respectively).

In this context, misreference or omission errors are easy to commit and hard to
detect.  Hence it is of some practical importance to organize the meta-logical
propositions attached to a particular reusable program part clearly and
accountably.  Adequate solutions appear to depend heavily on the programming
style; the following guidelines should thus be understood as both flexible and
incomplete.

\subsubsection{Theory Type Classes}

A substantial part of model-ish functional programs is about the algebra of data
structures.  For structures organized in the \REV{idiomatic} Haskell way as type
classes, the associated meta-logic can conveniently be organized as a companion
type class with default implementations.  This bundles the laws and makes them
accessible to \REV{simultaneous} instantiation, and to automatic enumeration via
meta-programming (which is not discussed here).

For example, consider the implied laws of the Prelude type class \ensuremath{\Conid{Monoid}}:
\begin{hscode}\SaveRestoreHook
\column{B}{@{}>{\hspre}l<{\hspost}@{}}%
\column{3}{@{}>{\hspre}l<{\hspost}@{}}%
\column{5}{@{}>{\hspre}l<{\hspost}@{}}%
\column{38}{@{}>{\hspre}l<{\hspost}@{}}%
\column{39}{@{}>{\hspre}l<{\hspost}@{}}%
\column{43}{@{}>{\hspre}l<{\hspost}@{}}%
\column{51}{@{}>{\hspre}c<{\hspost}@{}}%
\column{51E}{@{}l@{}}%
\column{52}{@{}>{\hspre}c<{\hspost}@{}}%
\column{52E}{@{}l@{}}%
\column{55}{@{}>{\hspre}l<{\hspost}@{}}%
\column{56}{@{}>{\hspre}l<{\hspost}@{}}%
\column{58}{@{}>{\hspre}c<{\hspost}@{}}%
\column{58E}{@{}l@{}}%
\column{62}{@{}>{\hspre}l<{\hspost}@{}}%
\column{E}{@{}>{\hspre}l<{\hspost}@{}}%
\>[3]{}\mathbf{class}\;(\Conid{Monoid}\;\alpha)\Rightarrow \Conid{MonoidTheory}\;\alpha\;\mathbf{where}{}\<[E]%
\\
\>[3]{}\hsindent{2}{}\<[5]%
\>[5]{}\Varid{monoid\char95 left\char95 unit}\mathbin{::}(\Conid{Eq}\;\alpha)\Rightarrow \Conid{Meta}\;(\alpha\to \Conid{Bool}){}\<[E]%
\\
\>[3]{}\hsindent{2}{}\<[5]%
\>[5]{}\Varid{monoid\char95 left\char95 unit}\mathrel{=}\Conid{Meta}\;(\lambda \Varid{x}\to {}\<[38]%
\>[38]{}\Varid{mempty}\mathbin{\diamond}\Varid{x}{}\<[51]%
\>[51]{}\equiv {}\<[51E]%
\>[55]{}\Varid{x}){}\<[E]%
\\[\blanklineskip]%
\>[3]{}\hsindent{2}{}\<[5]%
\>[5]{}\Varid{monoid\char95 right\char95 unit}\mathbin{::}(\Conid{Eq}\;\alpha)\Rightarrow \Conid{Meta}\;(\alpha\to \Conid{Bool}){}\<[E]%
\\
\>[3]{}\hsindent{2}{}\<[5]%
\>[5]{}\Varid{monoid\char95 right\char95 unit}\mathrel{=}\Conid{Meta}\;(\lambda \Varid{x}\to {}\<[39]%
\>[39]{}\Varid{x}\mathbin{\diamond}\Varid{mempty}{}\<[52]%
\>[52]{}\equiv {}\<[52E]%
\>[56]{}\Varid{x}){}\<[E]%
\\[\blanklineskip]%
\>[3]{}\hsindent{2}{}\<[5]%
\>[5]{}\Varid{monoid\char95 assoc}\mathbin{::}(\Conid{Eq}\;\alpha)\Rightarrow \Conid{Meta}\;((\alpha,\alpha,\alpha)\to \Conid{Bool}){}\<[E]%
\\
\>[3]{}\hsindent{2}{}\<[5]%
\>[5]{}\Varid{monoid\char95 assoc}\mathrel{=}\Conid{Meta}\;(\lambda (\Varid{x},\Varid{y},\Varid{z}){}\<[39]%
\>[39]{}\to {}\<[43]%
\>[43]{}(\Varid{x}\mathbin{\diamond}\Varid{y})\mathbin{\diamond}\Varid{z}{}\<[58]%
\>[58]{}\equiv {}\<[58E]%
\>[62]{}\Varid{x}\mathbin{\diamond}(\Varid{y}\mathbin{\diamond}\Varid{z})){}\<[E]%
\ColumnHook
\end{hscode}\resethooks
Note that there is some design leeway with respect to type class contexts.  For
illustration, we have distinguished here between the ``essential'' context
\ensuremath{\Conid{Monoid}\;\alpha}, declared on the type class and hence detected upon
instantiation, and the ``accidental'' context \ensuremath{\Conid{Eq}\;\alpha}, declared on each
method and hence detected upon use.  The distinction may or may not be ambiguous
in practice, however.

\subsubsection{Type-Level Ad-Hoc Programming}

For more ad-hoc data structures, where operations are not organized as methods
of a type class, but rather passed explicitly to higher-order functions, or
where extra laws are assumed locally, a likewise looser style of meta-logic
appears more adequate.  Fortunately, there is no need to relinquish the
assistance of the Haskell type and context checker altogether.  A type class can
be used to map symbolic names of laws, defined as constructors of ad-hoc
datatypes, to their logical content.
\begin{hscode}\SaveRestoreHook
\column{B}{@{}>{\hspre}l<{\hspost}@{}}%
\column{3}{@{}>{\hspre}l<{\hspost}@{}}%
\column{E}{@{}>{\hspre}l<{\hspost}@{}}%
\>[3]{}\mathbf{class}\;\Conid{Axiom}\;\alpha\;\pi\mid \alpha\to \pi\;\mathbf{where}\;\Varid{axiomatic}\mathbin{::}\alpha\to \Conid{Meta}\;\pi{}\<[E]%
\ColumnHook
\end{hscode}\resethooks
In line with the previous example, an extra law that is not reflected by a
Haskell type class can be defined and made referable by a singleton polymorphic
datatype.
\begin{hscode}\SaveRestoreHook
\column{B}{@{}>{\hspre}l<{\hspost}@{}}%
\column{3}{@{}>{\hspre}l<{\hspost}@{}}%
\column{5}{@{}>{\hspre}l<{\hspost}@{}}%
\column{13}{@{}>{\hspre}l<{\hspost}@{}}%
\column{E}{@{}>{\hspre}l<{\hspost}@{}}%
\>[3]{}\mathbf{data}\;\Conid{MonoidCommute}\;\alpha\mathrel{=}\Conid{MonoidCommute}{}\<[E]%
\\[\blanklineskip]%
\>[3]{}\mathbf{instance}\;{}\<[13]%
\>[13]{}(\Conid{Monoid}\;\alpha,\Conid{Eq}\;\alpha)\Rightarrow {}\<[E]%
\\
\>[13]{}\Conid{Axiom}\;(\Conid{MonoidCommute}\;\alpha)\;((\alpha,\alpha)\to \Conid{Bool})\;\mathbf{where}{}\<[E]%
\\
\>[3]{}\hsindent{2}{}\<[5]%
\>[5]{}\Varid{axiomatic}\;\Conid{MonoidCommute}\mathrel{=}\Conid{Meta}\;(\lambda (\Varid{x},\Varid{y})\to \Varid{x}\mathbin{\diamond}\Varid{y}\equiv \Varid{y}\mathbin{\diamond}\Varid{x}){}\<[E]%
\ColumnHook
\end{hscode}\resethooks
A law for an ad-hoc data structure with explicitly passed operations is
analogously defined as a record-like datatype. For instance consider a law of monoid
actions (also cf.\ the type signature of \ensuremath{\Varid{foldl}}):
\begin{hscode}\SaveRestoreHook
\column{B}{@{}>{\hspre}l<{\hspost}@{}}%
\column{3}{@{}>{\hspre}l<{\hspost}@{}}%
\column{5}{@{}>{\hspre}l<{\hspost}@{}}%
\column{13}{@{}>{\hspre}l<{\hspost}@{}}%
\column{E}{@{}>{\hspre}l<{\hspost}@{}}%
\>[3]{}\mathbf{type}\;\Conid{RAction}\;\alpha\;\beta\mathrel{=}\beta\to \alpha\to \beta{}\<[E]%
\\[\blanklineskip]%
\>[3]{}\mathbf{data}\;\Conid{RActionUnit}\;\alpha\;\beta\mathrel{=}\Conid{RActionUnit}\;(\Conid{RAction}\;\alpha\;\beta){}\<[E]%
\\[\blanklineskip]%
\>[3]{}\mathbf{instance}\;{}\<[13]%
\>[13]{}(\Conid{Monoid}\;\alpha,\Conid{Eq}\;\beta)\Rightarrow {}\<[E]%
\\
\>[13]{}\Conid{Axiom}\;(\Conid{RActionUnit}\;\alpha\;\beta)\;(\beta\to \Conid{Bool})\;\mathbf{where}{}\<[E]%
\\
\>[3]{}\hsindent{2}{}\<[5]%
\>[5]{}\Varid{axiomatic}\;(\Conid{RActionUnit}\;(\mathbin{\triangleleft}))\mathrel{=}\Conid{Meta}\;(\lambda \Varid{x}\to \Varid{x}\mathbin{\triangleleft}\Varid{mempty}\equiv \Varid{x}){}\<[E]%
\\[\blanklineskip]%
\>[3]{}\mathbf{data}\;\Conid{RActionCompose}\;\alpha\;\beta\mathrel{=}\Conid{RActionCompose}\;(\Conid{RAction}\;\alpha\;\beta){}\<[E]%
\\[\blanklineskip]%
\>[3]{}\mathbf{instance}\;{}\<[13]%
\>[13]{}(\Conid{Monoid}\;\alpha,\Conid{Eq}\;\beta)\Rightarrow {}\<[E]%
\\
\>[13]{}\Conid{Axiom}\;(\Conid{RActionCompose}\;\alpha\;\beta)\;(\beta\to \alpha\to \alpha\to \Conid{Bool})\;\mathbf{where}{}\<[E]%
\\
\>[3]{}\hsindent{2}{}\<[5]%
\>[5]{}\Varid{axiomatic}\;(\Conid{RActionCompose}\;(\mathbin{\triangleleft}))\mathrel{=}\Conid{Meta}\;(\lambda \Varid{x}\;\Varid{y}\;\Varid{z}\to \Varid{x}\mathbin{\triangleleft}(\Varid{y}\mathbin{\diamond}\Varid{z})\equiv (\Varid{x}\mathbin{\triangleleft}\Varid{y})\mathbin{\triangleleft}\Varid{z}){}\<[E]%
\ColumnHook
\end{hscode}\resethooks

\begin{REVENV}

For richer classification, type subclasses can be used to create ad-hoc subsets
of ``axiom space''.  This both adds to the documentation value of actual
meta-level code, and protects against misuse of tactics.  For instance, consider
a class of \ensuremath{\Conid{Int}}-parameterized meta-level expressions that need only be checked
for non-negative parameter values:
\begin{hscode}\SaveRestoreHook
\column{B}{@{}>{\hspre}l<{\hspost}@{}}%
\column{3}{@{}>{\hspre}l<{\hspost}@{}}%
\column{E}{@{}>{\hspre}l<{\hspost}@{}}%
\>[3]{}\mathbf{class}\;(\Conid{Axiom}\;\alpha\;(\Conid{Int}\to \beta))\Rightarrow \Conid{NonNegAxiom}\;\alpha\;\beta{}\<[E]%
\ColumnHook
\end{hscode}\resethooks
This subclass can be accompanied with an axiom-level operator, by giving a
constructor type and corresponding operational lifting:
\begin{hscode}\SaveRestoreHook
\column{B}{@{}>{\hspre}l<{\hspost}@{}}%
\column{3}{@{}>{\hspre}l<{\hspost}@{}}%
\column{5}{@{}>{\hspre}l<{\hspost}@{}}%
\column{E}{@{}>{\hspre}l<{\hspost}@{}}%
\>[3]{}\mathbf{data}\;\Conid{NonNeg}\;\alpha\mathrel{=}\Conid{NonNeg}\;\alpha{}\<[E]%
\\[\blanklineskip]%
\>[3]{}\mathbf{instance}\;(\Conid{NonNegAxiom}\;\alpha\;\beta)\Rightarrow \Conid{Axiom}\;(\Conid{NonNeg}\;\alpha)\;(\Conid{Int}\to \beta)\;\mathbf{where}{}\<[E]%
\\
\>[3]{}\hsindent{2}{}\<[5]%
\>[5]{}\Varid{axiomatic}\;(\Conid{NonNeg}\;\Varid{a})\mathrel{=}\Conid{Meta}\;(\Varid{reflect}\;(\Varid{axiomatic}\;\Varid{a})\mathbin{\circ}\Varid{abs}){}\<[E]%
\ColumnHook
\end{hscode}\resethooks
The restriction of the instance context to the subclass \ensuremath{\Conid{NonNegAxiom}} ensures
application of this (generally unsafe) tactic only to axioms that have an
explicit membership declaration, which can serve as an anchor for individual
justification, be it prose reasoning or checkable lemmata.
\end{REVENV}

\subsection{Constructive Existentials}

The natural logical reading of the type operator \ensuremath{(\to )} is universal
quantification.  But existential quantification also often arises in formulas,
either explicitly or by DeMorgan's laws, when universal quantification occurs in
a negative position, such as under negation or on the left hand side of
implication.

Checking existential quantification with the same sampling-based mechanisms as
universal quantification would break the monotonicity of heuristics\REV{:}  For
universal quantifiers, only false positives can arise if counterexamples exist
but are not present in the sample.  As such, confidence can only improve when
the sample size is increased.  By contrast, for existential quantifiers, false
negatives can arise when witness exist but are not present in the sample.  False
negatives are at best annoying when they occur at the top level and raise false
alarms, but at worst, when arising negatively nested in a complex formula, they
can make overall confidence \emph{decrease} with increasing sample size.

Therefore we propose to treat existential quantification as entirely distinct,
and in the true spirit of constructive logic, \REV{by effective Skolemization}.  To make an existential assertion
checkable, a witness must be provided in an effectively computable fashion.
\begin{hscode}\SaveRestoreHook
\column{B}{@{}>{\hspre}l<{\hspost}@{}}%
\column{3}{@{}>{\hspre}l<{\hspost}@{}}%
\column{E}{@{}>{\hspre}l<{\hspost}@{}}%
\>[3]{}\mathbf{class}\;\Conid{Witness}\;\alpha\;\beta\mid \alpha\to \beta\;\mathbf{where}\;\Varid{witness}\mathbin{::}\alpha\to \Conid{Maybe}\;\beta{}\<[E]%
\ColumnHook
\end{hscode}\resethooks
Here \ensuremath{\alpha} is a data type that encodes the meta-logical predicate to quantify,
and \ensuremath{\beta} is the domain to quantify over.  The ad-hoc polymorphic operation
\ensuremath{\Varid{witness}} may yield \ensuremath{\Conid{Nothing}} to indicate that no witness could be found for the
given predicate instance.  The extraction of a witness can then be composed with
a payload predicate to form bounded existential quantifications.
\begin{hscode}\SaveRestoreHook
\column{B}{@{}>{\hspre}l<{\hspost}@{}}%
\column{3}{@{}>{\hspre}l<{\hspost}@{}}%
\column{5}{@{}>{\hspre}l<{\hspost}@{}}%
\column{14}{@{}>{\hspre}c<{\hspost}@{}}%
\column{14E}{@{}l@{}}%
\column{18}{@{}>{\hspre}l<{\hspost}@{}}%
\column{E}{@{}>{\hspre}l<{\hspost}@{}}%
\>[3]{}\Varid{exists}\mathbin{::}(\Conid{Witness}\;\alpha\;\beta)\Rightarrow \alpha\to (\beta\to \Conid{Bool})\to \Conid{Bool}{}\<[E]%
\\
\>[3]{}\Varid{exists}\;\Varid{p}\;\Varid{q}\mathrel{=}\mathbf{case}\;\Varid{witness}\;\Varid{p}\;\mathbf{of}{}\<[E]%
\\
\>[3]{}\hsindent{2}{}\<[5]%
\>[5]{}\Conid{Just}\;\Varid{x}{}\<[14]%
\>[14]{}\to {}\<[14E]%
\>[18]{}\Varid{q}\;\Varid{x}{}\<[E]%
\\
\>[3]{}\hsindent{2}{}\<[5]%
\>[5]{}\Conid{Nothing}{}\<[14]%
\>[14]{}\to {}\<[14E]%
\>[18]{}\Conid{False}{}\<[E]%
\ColumnHook
\end{hscode}\resethooks
Note that the \ensuremath{\Varid{exists}} itself quantifier is not marked with \ensuremath{\Conid{Meta}}, as it is
perfectly suitable for use in the operational codebase layer as well.
\begin{hscode}\SaveRestoreHook
\column{B}{@{}>{\hspre}l<{\hspost}@{}}%
\column{3}{@{}>{\hspre}l<{\hspost}@{}}%
\column{5}{@{}>{\hspre}l<{\hspost}@{}}%
\column{14}{@{}>{\hspre}c<{\hspost}@{}}%
\column{14E}{@{}l@{}}%
\column{18}{@{}>{\hspre}l<{\hspost}@{}}%
\column{E}{@{}>{\hspre}l<{\hspost}@{}}%
\>[3]{}\Varid{existsSome}\mathbin{::}(\Conid{Witness}\;\alpha\;\beta)\Rightarrow \alpha\to \Conid{Bool}{}\<[E]%
\\
\>[3]{}\Varid{existsSome}\;\Varid{p}\mathrel{=}\Varid{exists}\;\Varid{p}\;(\Varid{const}\;\Conid{True}){}\<[E]%
\\[\blanklineskip]%
\>[3]{}\Varid{existsOrVacuous}\mathbin{::}(\Conid{Witness}\;\alpha\;\beta)\Rightarrow \alpha\to (\beta\to \Conid{Bool})\to \Conid{Bool}{}\<[E]%
\\
\>[3]{}\Varid{existsOrVacuous}\;\Varid{p}\;\Varid{q}\mathrel{=}\mathbf{case}\;\Varid{witness}\;\Varid{p}\;\mathbf{of}{}\<[E]%
\\
\>[3]{}\hsindent{2}{}\<[5]%
\>[5]{}\Conid{Just}\;\Varid{x}{}\<[14]%
\>[14]{}\to {}\<[14E]%
\>[18]{}\Varid{q}\;\Varid{x}{}\<[E]%
\\
\>[3]{}\hsindent{2}{}\<[5]%
\>[5]{}\Conid{Nothing}{}\<[14]%
\>[14]{}\to {}\<[14E]%
\>[18]{}\Conid{True}{}\<[E]%
\ColumnHook
\end{hscode}\resethooks

\section{Example Application: Theory of (String) Patches}

We illustrate the use and impact of the checking idioms described above by
applying them to a conceptual problem arising from real-world software
engineering research: An algebraic theory of compositional patching.

The generic level of the theory studies non-Abelian groups of patches acting
partially on some arbitrary state space.  As a simple but illuminating example
instance, we consider the particular space of ordinary character strings, and a
group generated by atomic \emph{insert} and \emph{delete} operations and their
evident semantics.  Establishing the decidability of the word problem of this
group is already a non-trivial modeling task, where the expressivity gained by
our proposed checking idioms comes in handy for rapid validation.

\subsection{Group Words and Actions}

The theoretical background for a type of patches \ensuremath{\pi} is its (right) action, a
partial function on some state space \ensuremath{\sigma}.
\begin{hscode}\SaveRestoreHook
\column{B}{@{}>{\hspre}l<{\hspost}@{}}%
\column{3}{@{}>{\hspre}l<{\hspost}@{}}%
\column{E}{@{}>{\hspre}l<{\hspost}@{}}%
\>[3]{}\mathbf{class}\;\Conid{Patch}\;\sigma\;\pi\;\mathbf{where}\;\Varid{action}\mathbin{::}\sigma\to \pi\to \Conid{Maybe}\;\sigma{}\<[E]%
\ColumnHook
\end{hscode}\resethooks
Application of patches can also be reverted.
\begin{hscode}\SaveRestoreHook
\column{B}{@{}>{\hspre}l<{\hspost}@{}}%
\column{3}{@{}>{\hspre}l<{\hspost}@{}}%
\column{E}{@{}>{\hspre}l<{\hspost}@{}}%
\>[3]{}\mathbf{class}\;(\Conid{Patch}\;\sigma\;\pi)\Rightarrow \Conid{InvPatch}\;\sigma\;\pi\;\mathbf{where}\;\Varid{undo}\mathbin{::}\sigma\to \pi\to \Conid{Maybe}\;\sigma{}\<[E]%
\ColumnHook
\end{hscode}\resethooks
This should be an inverse operation where defined:
\begin{hscode}\SaveRestoreHook
\column{B}{@{}>{\hspre}l<{\hspost}@{}}%
\column{3}{@{}>{\hspre}l<{\hspost}@{}}%
\column{5}{@{}>{\hspre}l<{\hspost}@{}}%
\column{13}{@{}>{\hspre}l<{\hspost}@{}}%
\column{45}{@{}>{\hspre}l<{\hspost}@{}}%
\column{47}{@{}>{\hspre}l<{\hspost}@{}}%
\column{56}{@{}>{\hspre}c<{\hspost}@{}}%
\column{56E}{@{}l@{}}%
\column{60}{@{}>{\hspre}l<{\hspost}@{}}%
\column{E}{@{}>{\hspre}l<{\hspost}@{}}%
\>[3]{}\mathbf{data}\;\Conid{PatchInvert}\;\sigma\;\pi\mathrel{=}\Conid{PatchInvert}{}\<[E]%
\\[\blanklineskip]%
\>[3]{}\mathbf{instance}\;{}\<[13]%
\>[13]{}(\Conid{InvPatch}\;\sigma\;\pi,\Conid{Eq}\;\sigma)\Rightarrow {}\<[E]%
\\
\>[13]{}\Conid{Axiom}\;(\Conid{PatchInvert}\;\sigma\;\pi)\;(\sigma\to \pi\to \Conid{Bool})\;\mathbf{where}{}\<[E]%
\\
\>[3]{}\hsindent{2}{}\<[5]%
\>[5]{}\Varid{axiomatic}\;\Conid{PatchInvert}\mathrel{=}\Conid{Meta}\;(\lambda \Varid{s}\;\Varid{p}\to {}\<[45]%
\>[45]{}\mathbf{case}\;\Varid{action}\;\Varid{s}\;\Varid{p}\;\mathbf{of}{}\<[E]%
\\
\>[45]{}\hsindent{2}{}\<[47]%
\>[47]{}\Conid{Nothing}{}\<[56]%
\>[56]{}\to {}\<[56E]%
\>[60]{}\Conid{True}{}\<[E]%
\\
\>[45]{}\hsindent{2}{}\<[47]%
\>[47]{}\Conid{Just}\;\Varid{s'}{}\<[56]%
\>[56]{}\to {}\<[56E]%
\>[60]{}\Varid{undo}\;\Varid{s'}\;\Varid{p}\equiv \Conid{Just}\;\Varid{s}){}\<[E]%
\ColumnHook
\end{hscode}\resethooks

If the patch type has a \emph{polarity}, that is some internal form of
inversion, then the forward direction \ensuremath{\Varid{action}} suffices to imply the backward
direction \ensuremath{\Varid{undo}}.
\begin{hscode}\SaveRestoreHook
\column{B}{@{}>{\hspre}l<{\hspost}@{}}%
\column{3}{@{}>{\hspre}l<{\hspost}@{}}%
\column{5}{@{}>{\hspre}l<{\hspost}@{}}%
\column{15}{@{}>{\hspre}c<{\hspost}@{}}%
\column{15E}{@{}l@{}}%
\column{18}{@{}>{\hspre}l<{\hspost}@{}}%
\column{E}{@{}>{\hspre}l<{\hspost}@{}}%
\>[3]{}\mathbf{class}\;\Conid{Polar}\;\alpha\;\mathbf{where}{}\<[E]%
\\
\>[3]{}\hsindent{2}{}\<[5]%
\>[5]{}\Varid{inv}\mathbin{::}\alpha\to \alpha{}\<[E]%
\\[\blanklineskip]%
\>[3]{}\mathbf{instance}\;(\Conid{Patch}\;\sigma\;\pi,\Conid{Polar}\;\pi)\Rightarrow \Conid{InvPatch}\;\sigma\;\pi\;\mathbf{where}{}\<[E]%
\\
\>[3]{}\hsindent{2}{}\<[5]%
\>[5]{}\Varid{undo}\;\Varid{x}\;\Varid{p}{}\<[15]%
\>[15]{}\mathrel{=}{}\<[15E]%
\>[18]{}\Varid{action}\;\Varid{x}\;(\Varid{inv}\;\Varid{p}){}\<[E]%
\ColumnHook
\end{hscode}\resethooks

The most important forms of patch types are \emph{group words}, made up from
polarized primitives:
\begin{hscode}\SaveRestoreHook
\column{B}{@{}>{\hspre}l<{\hspost}@{}}%
\column{3}{@{}>{\hspre}l<{\hspost}@{}}%
\column{5}{@{}>{\hspre}l<{\hspost}@{}}%
\column{19}{@{}>{\hspre}c<{\hspost}@{}}%
\column{19E}{@{}l@{}}%
\column{22}{@{}>{\hspre}l<{\hspost}@{}}%
\column{E}{@{}>{\hspre}l<{\hspost}@{}}%
\>[3]{}\mathbf{data}\;\Conid{Polarity}\mathrel{=}\Conid{Positive}\mid \Conid{Negative}{}\<[E]%
\\[\blanklineskip]%
\>[3]{}\mathbf{instance}\;\Conid{Polar}\;\Conid{Polarity}\;\mathbf{where}{}\<[E]%
\\
\>[3]{}\hsindent{2}{}\<[5]%
\>[5]{}\Varid{inv}\;\Conid{Positive}{}\<[19]%
\>[19]{}\mathrel{=}{}\<[19E]%
\>[22]{}\Conid{Negative}{}\<[E]%
\\
\>[3]{}\hsindent{2}{}\<[5]%
\>[5]{}\Varid{inv}\;\Conid{Negative}{}\<[19]%
\>[19]{}\mathrel{=}{}\<[19E]%
\>[22]{}\Conid{Positive}{}\<[E]%
\ColumnHook
\end{hscode}\resethooks
\begin{hscode}\SaveRestoreHook
\column{B}{@{}>{\hspre}l<{\hspost}@{}}%
\column{3}{@{}>{\hspre}l<{\hspost}@{}}%
\column{5}{@{}>{\hspre}l<{\hspost}@{}}%
\column{23}{@{}>{\hspre}c<{\hspost}@{}}%
\column{23E}{@{}l@{}}%
\column{24}{@{}>{\hspre}c<{\hspost}@{}}%
\column{24E}{@{}l@{}}%
\column{26}{@{}>{\hspre}l<{\hspost}@{}}%
\column{27}{@{}>{\hspre}l<{\hspost}@{}}%
\column{33}{@{}>{\hspre}l<{\hspost}@{}}%
\column{37}{@{}>{\hspre}c<{\hspost}@{}}%
\column{37E}{@{}l@{}}%
\column{40}{@{}>{\hspre}l<{\hspost}@{}}%
\column{48}{@{}>{\hspre}l<{\hspost}@{}}%
\column{E}{@{}>{\hspre}l<{\hspost}@{}}%
\>[3]{}\mathbf{data}\;\Conid{Literal}\;\alpha{}\<[23]%
\>[23]{}\mathrel{=}{}\<[23E]%
\>[26]{}\Conid{Literal}\;\Conid{Polarity}\;\alpha{}\<[E]%
\\[\blanklineskip]%
\>[3]{}\mathbf{instance}\;\Conid{Polar}\;(\Conid{Literal}\;\alpha)\;\mathbf{where}{}\<[E]%
\\
\>[3]{}\hsindent{2}{}\<[5]%
\>[5]{}\Varid{inv}\;(\Conid{Literal}\;\Varid{b}\;\Varid{x}){}\<[24]%
\>[24]{}\mathrel{=}{}\<[24E]%
\>[27]{}\Conid{Literal}\;(\Varid{inv}\;\Varid{b})\;\Varid{x}{}\<[E]%
\\[\blanklineskip]%
\>[3]{}\mathbf{instance}\;(\Conid{InvPatch}\;\sigma\;\alpha)\Rightarrow \Conid{Patch}\;\sigma\;(\Conid{Literal}\;\alpha)\;\mathbf{where}{}\<[E]%
\\
\>[3]{}\hsindent{2}{}\<[5]%
\>[5]{}\Varid{action}\;\Varid{s}\;(\Conid{Literal}\;\Conid{Positive}\;{}\<[33]%
\>[33]{}\Varid{p}){}\<[37]%
\>[37]{}\mathrel{=}{}\<[37E]%
\>[40]{}\Varid{action}\;{}\<[48]%
\>[48]{}\Varid{s}\;\Varid{p}{}\<[E]%
\\
\>[3]{}\hsindent{2}{}\<[5]%
\>[5]{}\Varid{action}\;\Varid{s}\;(\Conid{Literal}\;\Conid{Negative}\;{}\<[33]%
\>[33]{}\Varid{p}){}\<[37]%
\>[37]{}\mathrel{=}{}\<[37E]%
\>[40]{}\Varid{undo}\;{}\<[48]%
\>[48]{}\Varid{s}\;\Varid{p}{}\<[E]%
\ColumnHook
\end{hscode}\resethooks
Group words are essentially lists that polarize elementwise, but also reverse
their order in the process, to accomodate for non-commutative groups.
\begin{hscode}\SaveRestoreHook
\column{B}{@{}>{\hspre}l<{\hspost}@{}}%
\column{3}{@{}>{\hspre}l<{\hspost}@{}}%
\column{5}{@{}>{\hspre}l<{\hspost}@{}}%
\column{E}{@{}>{\hspre}l<{\hspost}@{}}%
\>[3]{}\mathbf{newtype}\;\Conid{Word}\;\alpha\mathrel{=}\Conid{Word}\;[\mskip1.5mu \Conid{Literal}\;\alpha\mskip1.5mu]{}\<[E]%
\\[\blanklineskip]%
\>[3]{}\mathbf{instance}\;\Conid{Polar}\;(\Conid{Word}\;\alpha)\;\mathbf{where}{}\<[E]%
\\
\>[3]{}\hsindent{2}{}\<[5]%
\>[5]{}\Varid{inv}\;(\Conid{Word}\;\Varid{w})\mathrel{=}\Conid{Word}\;(\Varid{reverse}\;(\Varid{map}\;\Varid{inv}\;\Varid{w})){}\<[E]%
\ColumnHook
\end{hscode}\resethooks
They act in the obvious way by folding, strictly over the \ensuremath{\Conid{Maybe}} monad.
\begin{hscode}\SaveRestoreHook
\column{B}{@{}>{\hspre}l<{\hspost}@{}}%
\column{3}{@{}>{\hspre}l<{\hspost}@{}}%
\column{5}{@{}>{\hspre}l<{\hspost}@{}}%
\column{24}{@{}>{\hspre}c<{\hspost}@{}}%
\column{24E}{@{}l@{}}%
\column{27}{@{}>{\hspre}l<{\hspost}@{}}%
\column{E}{@{}>{\hspre}l<{\hspost}@{}}%
\>[3]{}\mathbf{instance}\;(\Conid{InvPatch}\;\sigma\;\alpha)\Rightarrow \Conid{Patch}\;\sigma\;(\Conid{Word}\;\alpha)\;\mathbf{where}{}\<[E]%
\\
\>[3]{}\hsindent{2}{}\<[5]%
\>[5]{}\Varid{action}\;\Varid{s}\;(\Conid{Word}\;\Varid{w}){}\<[24]%
\>[24]{}\mathrel{=}{}\<[24E]%
\>[27]{}\Varid{foldM}\;\Varid{action}\;\Varid{s}\;\Varid{w}{}\<[E]%
\ColumnHook
\end{hscode}\resethooks
\subsection{String Editing Operations}

As an example instance of the generic theory, consider the editing of a
character string.  Suitable partial invertible atomic edit operations are:

\begin{itemize}
\item Inserting a given character at a given position if that does not exceed the end of the string, and inversely
\item deleting a given character at a given position if it occurs there.
\end{itemize}
\begin{hscode}\SaveRestoreHook
\column{B}{@{}>{\hspre}l<{\hspost}@{}}%
\column{3}{@{}>{\hspre}l<{\hspost}@{}}%
\column{5}{@{}>{\hspre}l<{\hspost}@{}}%
\column{17}{@{}>{\hspre}c<{\hspost}@{}}%
\column{17E}{@{}l@{}}%
\column{20}{@{}>{\hspre}l<{\hspost}@{}}%
\column{E}{@{}>{\hspre}l<{\hspost}@{}}%
\>[3]{}\mathbf{data}\;\Conid{EditOp}\mathrel{=}\Conid{Insert}\mid \Conid{Delete}{}\<[E]%
\\[\blanklineskip]%
\>[3]{}\mathbf{instance}\;\Conid{Polar}\;\Conid{EditOp}\;\mathbf{where}{}\<[E]%
\\
\>[3]{}\hsindent{2}{}\<[5]%
\>[5]{}\Varid{inv}\;\Conid{Insert}{}\<[17]%
\>[17]{}\mathrel{=}{}\<[17E]%
\>[20]{}\Conid{Delete}{}\<[E]%
\\
\>[3]{}\hsindent{2}{}\<[5]%
\>[5]{}\Varid{inv}\;\Conid{Delete}{}\<[17]%
\>[17]{}\mathrel{=}{}\<[17E]%
\>[20]{}\Conid{Insert}{}\<[E]%
\ColumnHook
\end{hscode}\resethooks
\begin{hscode}\SaveRestoreHook
\column{B}{@{}>{\hspre}l<{\hspost}@{}}%
\column{3}{@{}>{\hspre}l<{\hspost}@{}}%
\column{5}{@{}>{\hspre}l<{\hspost}@{}}%
\column{14}{@{}>{\hspre}c<{\hspost}@{}}%
\column{14E}{@{}l@{}}%
\column{17}{@{}>{\hspre}l<{\hspost}@{}}%
\column{23}{@{}>{\hspre}c<{\hspost}@{}}%
\column{23E}{@{}l@{}}%
\column{26}{@{}>{\hspre}l<{\hspost}@{}}%
\column{E}{@{}>{\hspre}l<{\hspost}@{}}%
\>[3]{}\mathbf{data}\;\Conid{Edit}{}\<[14]%
\>[14]{}\mathrel{=}{}\<[14E]%
\>[17]{}\Conid{Edit}\;\{\mskip1.5mu \Varid{op}\mathbin{::}\Conid{EditOp},\Varid{pos}\mathbin{::}\Conid{Int},\Varid{arg}\mathbin{::}\Conid{Char}\mskip1.5mu\}{}\<[E]%
\\[\blanklineskip]%
\>[3]{}\mathbf{instance}\;\Conid{Polar}\;\Conid{Edit}\;\mathbf{where}{}\<[E]%
\\
\>[3]{}\hsindent{2}{}\<[5]%
\>[5]{}\Varid{inv}\;(\Conid{Edit}\;\Varid{f}\;\Varid{i}\;\Varid{x}){}\<[23]%
\>[23]{}\mathrel{=}{}\<[23E]%
\>[26]{}\Conid{Edit}\;(\Varid{inv}\;\Varid{f})\;\Varid{i}\;\Varid{x}{}\<[E]%
\ColumnHook
\end{hscode}\resethooks
The operational semantics are modeled effectively by a type class that
interprets the two operations, giving rise to an action on some state space.
\begin{hscode}\SaveRestoreHook
\column{B}{@{}>{\hspre}l<{\hspost}@{}}%
\column{3}{@{}>{\hspre}l<{\hspost}@{}}%
\column{5}{@{}>{\hspre}l<{\hspost}@{}}%
\column{13}{@{}>{\hspre}c<{\hspost}@{}}%
\column{13E}{@{}l@{}}%
\column{17}{@{}>{\hspre}l<{\hspost}@{}}%
\column{21}{@{}>{\hspre}l<{\hspost}@{}}%
\column{29}{@{}>{\hspre}l<{\hspost}@{}}%
\column{35}{@{}>{\hspre}c<{\hspost}@{}}%
\column{35E}{@{}l@{}}%
\column{38}{@{}>{\hspre}l<{\hspost}@{}}%
\column{46}{@{}>{\hspre}l<{\hspost}@{}}%
\column{E}{@{}>{\hspre}l<{\hspost}@{}}%
\>[3]{}\mathbf{class}\;\Conid{Editable}\;\alpha\;\mathbf{where}{}\<[E]%
\\
\>[3]{}\hsindent{2}{}\<[5]%
\>[5]{}\Varid{insert}{}\<[13]%
\>[13]{}\mathbin{::}{}\<[13E]%
\>[17]{}\alpha\to \Conid{Int}\to \Conid{Char}\to \Conid{Maybe}\;\alpha{}\<[E]%
\\
\>[3]{}\hsindent{2}{}\<[5]%
\>[5]{}\Varid{delete}{}\<[13]%
\>[13]{}\mathbin{::}{}\<[13E]%
\>[17]{}\alpha\to \Conid{Int}\to \Conid{Char}\to \Conid{Maybe}\;\alpha{}\<[E]%
\\[\blanklineskip]%
\>[3]{}\mathbf{instance}\;(\Conid{Editable}\;\sigma)\Rightarrow \Conid{Patch}\;\sigma\;\Conid{Edit}\;\mathbf{where}{}\<[E]%
\\
\>[3]{}\hsindent{2}{}\<[5]%
\>[5]{}\Varid{action}\;\Varid{s}\;(\Conid{Edit}\;{}\<[21]%
\>[21]{}\Conid{Insert}\;{}\<[29]%
\>[29]{}\Varid{i}\;\Varid{x}){}\<[35]%
\>[35]{}\mathrel{=}{}\<[35E]%
\>[38]{}\Varid{insert}\;{}\<[46]%
\>[46]{}\Varid{s}\;\Varid{i}\;\Varid{x}{}\<[E]%
\\
\>[3]{}\hsindent{2}{}\<[5]%
\>[5]{}\Varid{action}\;\Varid{s}\;(\Conid{Edit}\;{}\<[21]%
\>[21]{}\Conid{Delete}\;{}\<[29]%
\>[29]{}\Varid{i}\;\Varid{x}){}\<[35]%
\>[35]{}\mathrel{=}{}\<[35E]%
\>[38]{}\Varid{delete}\;{}\<[46]%
\>[46]{}\Varid{s}\;\Varid{i}\;\Varid{x}{}\<[E]%
\ColumnHook
\end{hscode}\resethooks
The instance for the datatype \ensuremath{\Conid{String}} implements the above informal intuition.
\begin{hscode}\SaveRestoreHook
\column{B}{@{}>{\hspre}l<{\hspost}@{}}%
\column{3}{@{}>{\hspre}l<{\hspost}@{}}%
\column{5}{@{}>{\hspre}l<{\hspost}@{}}%
\column{13}{@{}>{\hspre}l<{\hspost}@{}}%
\column{22}{@{}>{\hspre}l<{\hspost}@{}}%
\column{25}{@{}>{\hspre}l<{\hspost}@{}}%
\column{28}{@{}>{\hspre}c<{\hspost}@{}}%
\column{28E}{@{}l@{}}%
\column{31}{@{}>{\hspre}l<{\hspost}@{}}%
\column{E}{@{}>{\hspre}l<{\hspost}@{}}%
\>[3]{}\mathbf{instance}\;\Conid{Editable}\;\Conid{String}\;\mathbf{where}{}\<[E]%
\\
\>[3]{}\hsindent{2}{}\<[5]%
\>[5]{}\Varid{insert}\;{}\<[13]%
\>[13]{}\Varid{s}\;{}\<[22]%
\>[22]{}\mathrm{0}\;{}\<[25]%
\>[25]{}\Varid{x}{}\<[28]%
\>[28]{}\mathrel{=}{}\<[28E]%
\>[31]{}\Varid{return}\;(\Varid{x}\mathbin{:}\Varid{s}){}\<[E]%
\\
\>[3]{}\hsindent{2}{}\<[5]%
\>[5]{}\Varid{insert}\;{}\<[13]%
\>[13]{}[\mskip1.5mu \mskip1.5mu]\;{}\<[22]%
\>[22]{}\Varid{i}\;{}\<[25]%
\>[25]{}\Varid{x}{}\<[28]%
\>[28]{}\mathrel{=}{}\<[28E]%
\>[31]{}\Conid{Nothing}{}\<[E]%
\\
\>[3]{}\hsindent{2}{}\<[5]%
\>[5]{}\Varid{insert}\;{}\<[13]%
\>[13]{}(\Varid{y}\mathbin{:}\Varid{t})\;{}\<[22]%
\>[22]{}\Varid{i}\;{}\<[25]%
\>[25]{}\Varid{x}{}\<[28]%
\>[28]{}\mathrel{=}{}\<[28E]%
\>[31]{}\Varid{liftM}\;(\Varid{y}\mathbin{:})\;(\Varid{insert}\;\Varid{t}\;(\Varid{i}\mathbin{-}\mathrm{1})\;\Varid{x}){}\<[E]%
\\[\blanklineskip]%
\>[3]{}\hsindent{2}{}\<[5]%
\>[5]{}\Varid{delete}\;{}\<[13]%
\>[13]{}[\mskip1.5mu \mskip1.5mu]\;{}\<[22]%
\>[22]{}\Varid{i}\;{}\<[25]%
\>[25]{}\Varid{x}{}\<[28]%
\>[28]{}\mathrel{=}{}\<[28E]%
\>[31]{}\Conid{Nothing}{}\<[E]%
\\
\>[3]{}\hsindent{2}{}\<[5]%
\>[5]{}\Varid{delete}\;{}\<[13]%
\>[13]{}(\Varid{y}\mathbin{:}\Varid{t})\;{}\<[22]%
\>[22]{}\mathrm{0}\;{}\<[25]%
\>[25]{}\Varid{x}{}\<[28]%
\>[28]{}\mathrel{=}{}\<[28E]%
\>[31]{}\Varid{guard}\;(\Varid{x}\equiv \Varid{y})\sequ \Varid{return}\;\Varid{t}{}\<[E]%
\\
\>[3]{}\hsindent{2}{}\<[5]%
\>[5]{}\Varid{delete}\;{}\<[13]%
\>[13]{}(\Varid{y}\mathbin{:}\Varid{t})\;{}\<[22]%
\>[22]{}\Varid{i}\;{}\<[25]%
\>[25]{}\Varid{x}{}\<[28]%
\>[28]{}\mathrel{=}{}\<[28E]%
\>[31]{}\Varid{liftM}\;(\Varid{y}\mathbin{:})\;(\Varid{delete}\;\Varid{t}\;(\Varid{i}\mathbin{-}\mathrm{1})\;\Varid{x}){}\<[E]%
\ColumnHook
\end{hscode}\resethooks

\subsection{Semantic Model}

Group words only form a free monoid, in the obvious way inherited from \REV{the list type} \ensuremath{[\mskip1.5mu \mskip1.5mu]}, but
partial applications of \ensuremath{\Varid{flip}\;\Varid{action}} induce a proper group of partial
bijections on the state space.  The extensional equality of induced group
elements, and thus the \emph{word problem} of the group presentation encoded in
the action, is universally quantified, and thus hard to decide for large or even
infinite state spaces.  A heuristic evaluation would be sufficient as a
meta-expression in simple offline checks, but not in negative positions, nor for
online assertions, nor even in the operational layer of the codebase, such as in
model animations.

This situation can be improved substantially by giving a semantic model in the
form of an algebraic datatype with inductively derived equality, which is
\emph{fully abstract} in the sense that it admits a normal form where
extensionally equal semantic functions are represented by the same data value.

For the example theory considered here, there is such a normal form of
string-transducing automata.  Because these automata do not require circular
transitions, they can be modeled by a family of mutually linearly recursive
datatypes, and evaluated by straightforward recursion.
\begin{hscode}\SaveRestoreHook
\column{B}{@{}>{\hspre}l<{\hspost}@{}}%
\column{3}{@{}>{\hspre}l<{\hspost}@{}}%
\column{21}{@{}>{\hspre}c<{\hspost}@{}}%
\column{21E}{@{}l@{}}%
\column{24}{@{}>{\hspre}l<{\hspost}@{}}%
\column{39}{@{}>{\hspre}c<{\hspost}@{}}%
\column{39E}{@{}l@{}}%
\column{42}{@{}>{\hspre}l<{\hspost}@{}}%
\column{E}{@{}>{\hspre}l<{\hspost}@{}}%
\>[3]{}\mathbf{data}\;\Conid{Editor}{}\<[21]%
\>[21]{}\mathrel{=}{}\<[21E]%
\>[24]{}\Conid{Try}\;\Conid{Insertion}{}\<[39]%
\>[39]{}\mid {}\<[39E]%
\>[42]{}\Conid{Fail}{}\<[E]%
\\[\blanklineskip]%
\>[3]{}\mathbf{data}\;\Conid{Insertion}{}\<[21]%
\>[21]{}\mathrel{=}{}\<[21E]%
\>[24]{}\Conid{Ins}\;\Conid{String}\;\Conid{Consumption}{}\<[E]%
\\[\blanklineskip]%
\>[3]{}\mathbf{data}\;\Conid{Consumption}{}\<[21]%
\>[21]{}\mathrel{=}{}\<[21E]%
\>[24]{}\Conid{Skip}\;\Conid{Insertion}{}\<[E]%
\\
\>[21]{}\mid {}\<[21E]%
\>[24]{}\Conid{Del}\;\Conid{Char}\;\Conid{Insertion}{}\<[E]%
\\
\>[21]{}\mid {}\<[21E]%
\>[24]{}\Conid{Return}{}\<[E]%
\ColumnHook
\end{hscode}\resethooks
The operational idea is to apply each operator node to a position in an input
string, advancing left to right.  The detailed meaning of operators is as
follows:

\begin{description}\small
\item[\ensuremath{\Conid{Fail}}] Applies to no string at all, immediately reject.
\item[\ensuremath{\Conid{Try}}] Applies to some strings, begin processing at start position.
\item[\ensuremath{\Conid{Ins}}] Insert zero or more characters before the position.
\item[\ensuremath{\Conid{Skip}}] Advance the position over one character if available, otherwise reject.
\item[\ensuremath{\Conid{Del}}] Remove the next character if available and matched, otherwise reject.
\item[\ensuremath{\Conid{Return}}] Stop processing and accept, returning the remainder of the string as is.
\end{description}

The sorting of operators into different data types ensures that insertion and
consumption alternate properly.

Note that, unlike random-access \ensuremath{\Conid{Edit}} terms, subsequent operator nodes are only
ever applied to the original input string, not to the output of their
predecessors.  This is also the cause for the \ensuremath{\Conid{Skip}} and \ensuremath{\Conid{Fail}} operators which
do not appear in the \ensuremath{\Conid{Edit}} language; they arise from attempting to delete a
previously inserted character consistently and inconsistently, respectively.

The type \ensuremath{\Conid{Insertion}} is not to be constructed directly, but by the following
smart constructor that avoids a degenerate corner case: Namely, insertions
before and after a deletion are operationally indistinguishable.  This ambiguity
is avoided by avoiding insertions after a deletion, lumping adjacent insertions
together beforehands, which preserves the desired normal form.
\begin{hscode}\SaveRestoreHook
\column{B}{@{}>{\hspre}l<{\hspost}@{}}%
\column{3}{@{}>{\hspre}l<{\hspost}@{}}%
\column{8}{@{}>{\hspre}c<{\hspost}@{}}%
\column{8E}{@{}l@{}}%
\column{12}{@{}>{\hspre}l<{\hspost}@{}}%
\column{20}{@{}>{\hspre}c<{\hspost}@{}}%
\column{20E}{@{}l@{}}%
\column{24}{@{}>{\hspre}l<{\hspost}@{}}%
\column{35}{@{}>{\hspre}c<{\hspost}@{}}%
\column{35E}{@{}l@{}}%
\column{37}{@{}>{\hspre}c<{\hspost}@{}}%
\column{37E}{@{}l@{}}%
\column{38}{@{}>{\hspre}l<{\hspost}@{}}%
\column{41}{@{}>{\hspre}l<{\hspost}@{}}%
\column{E}{@{}>{\hspre}l<{\hspost}@{}}%
\>[3]{}\Varid{ins}{}\<[8]%
\>[8]{}\mathbin{::}{}\<[8E]%
\>[12]{}\Conid{String}{}\<[20]%
\>[20]{}\to {}\<[20E]%
\>[24]{}\Conid{Consumption}{}\<[37]%
\>[37]{}\to {}\<[37E]%
\>[41]{}\Conid{Insertion}{}\<[E]%
\\
\>[3]{}\Varid{ins}\;\Varid{pre}\;(\Conid{Del}\;\Varid{y}\;(\Conid{Ins}\;\Varid{fix}\;\Varid{next})){}\<[35]%
\>[35]{}\mathrel{=}{}\<[35E]%
\>[38]{}\Conid{Ins}\;(\Varid{pre}\plus \Varid{fix})\;(\Conid{Del}\;\Varid{y}\;(\Conid{Ins}\;[\mskip1.5mu \mskip1.5mu]\;\Varid{next})){}\<[E]%
\\
\>[3]{}\Varid{ins}\;\Varid{prefix}\;\Varid{next}{}\<[35]%
\>[35]{}\mathrel{=}{}\<[35E]%
\>[38]{}\Conid{Ins}\;\Varid{prefix}\;\Varid{next}{}\<[E]%
\ColumnHook
\end{hscode}\resethooks

The semantic model type \ensuremath{\Conid{Editor}} covers the middle ground between the syntactic
encoding \ensuremath{\Conid{Word}\;\Conid{Edit}} and the semantic state space \ensuremath{\Conid{String}}, in the sense that it
instantiates both \ensuremath{\Conid{Editable}} and \ensuremath{\Conid{Patch}\;\Conid{String}}, giving the respective effective
connections.  The latter is the simpler one of the pair, and implements the
intuition stated above.
\begin{hscode}\SaveRestoreHook
\column{B}{@{}>{\hspre}l<{\hspost}@{}}%
\column{3}{@{}>{\hspre}l<{\hspost}@{}}%
\column{5}{@{}>{\hspre}l<{\hspost}@{}}%
\column{22}{@{}>{\hspre}l<{\hspost}@{}}%
\column{34}{@{}>{\hspre}c<{\hspost}@{}}%
\column{34E}{@{}l@{}}%
\column{37}{@{}>{\hspre}l<{\hspost}@{}}%
\column{41}{@{}>{\hspre}l<{\hspost}@{}}%
\column{49}{@{}>{\hspre}c<{\hspost}@{}}%
\column{49E}{@{}l@{}}%
\column{53}{@{}>{\hspre}l<{\hspost}@{}}%
\column{E}{@{}>{\hspre}l<{\hspost}@{}}%
\>[3]{}\mathbf{instance}\;\Conid{Patch}\;\Conid{String}\;\Conid{Editor}\;\mathbf{where}{}\<[E]%
\\
\>[3]{}\hsindent{2}{}\<[5]%
\>[5]{}\Varid{action}\;\Varid{s}\;\Conid{Fail}{}\<[34]%
\>[34]{}\mathrel{=}{}\<[34E]%
\>[37]{}\Conid{Nothing}{}\<[E]%
\\
\>[3]{}\hsindent{2}{}\<[5]%
\>[5]{}\Varid{action}\;\Varid{s}\;(\Conid{Try}\;\Varid{steps}){}\<[34]%
\>[34]{}\mathrel{=}{}\<[34E]%
\>[37]{}\Varid{action}\;\Varid{s}\;\Varid{steps}{}\<[E]%
\\[\blanklineskip]%
\>[3]{}\mathbf{instance}\;\Conid{Patch}\;\Conid{String}\;\Conid{Insertion}\;\mathbf{where}{}\<[E]%
\\
\>[3]{}\hsindent{2}{}\<[5]%
\>[5]{}\Varid{action}\;\Varid{s}\;(\Conid{Ins}\;\Varid{prefix}\;\Varid{next}){}\<[34]%
\>[34]{}\mathrel{=}{}\<[34E]%
\>[37]{}\mathbf{do}\;{}\<[41]%
\>[41]{}\Varid{t}\leftarrow \Varid{action}\;\Varid{s}\;\Varid{next}{}\<[E]%
\\
\>[41]{}\Varid{return}\;(\Varid{prefix}\plus \Varid{t}){}\<[E]%
\\[\blanklineskip]%
\>[3]{}\mathbf{instance}\;\Conid{Patch}\;\Conid{String}\;\Conid{Consumption}\;\mathbf{where}{}\<[E]%
\\
\>[3]{}\hsindent{2}{}\<[5]%
\>[5]{}\Varid{action}\;\Varid{s}\;\Conid{Return}{}\<[34]%
\>[34]{}\mathrel{=}{}\<[34E]%
\>[37]{}\Varid{return}\;\Varid{s}{}\<[E]%
\\
\>[3]{}\hsindent{2}{}\<[5]%
\>[5]{}\Varid{action}\;\Varid{s}\;(\Conid{Skip}\;{}\<[22]%
\>[22]{}\Varid{rest}){}\<[34]%
\>[34]{}\mathrel{=}{}\<[34E]%
\>[37]{}\mathbf{do}\;{}\<[41]%
\>[41]{}(\Varid{x},\Varid{t}){}\<[49]%
\>[49]{}\leftarrow {}\<[49E]%
\>[53]{}\Varid{uncons}\;\Varid{s}{}\<[E]%
\\
\>[41]{}\Varid{u}{}\<[49]%
\>[49]{}\leftarrow {}\<[49E]%
\>[53]{}\Varid{action}\;\Varid{t}\;\Varid{rest}{}\<[E]%
\\
\>[41]{}\Varid{return}\;(\Varid{x}\mathbin{:}\Varid{u}){}\<[E]%
\\
\>[3]{}\hsindent{2}{}\<[5]%
\>[5]{}\Varid{action}\;\Varid{s}\;(\Conid{Del}\;\Varid{y}\;{}\<[22]%
\>[22]{}\Varid{rest}){}\<[34]%
\>[34]{}\mathrel{=}{}\<[34E]%
\>[37]{}\mathbf{do}\;{}\<[41]%
\>[41]{}(\Varid{x},\Varid{t}){}\<[49]%
\>[49]{}\leftarrow {}\<[49E]%
\>[53]{}\Varid{uncons}\;\Varid{s}{}\<[E]%
\\
\>[41]{}\Varid{u}{}\<[49]%
\>[49]{}\leftarrow {}\<[49E]%
\>[53]{}\Varid{action}\;\Varid{t}\;\Varid{rest}{}\<[E]%
\\
\>[41]{}\Varid{guard}\;(\Varid{x}\equiv \Varid{y}){}\<[E]%
\\
\>[41]{}\Varid{return}\;\Varid{u}{}\<[E]%
\ColumnHook
\end{hscode}\resethooks

The instantiation of \ensuremath{\Conid{Editable}} \REV{essentially} amounts to splicing a single
edit operation into an automaton while preserving the normal form.  The
technical details are too gruesome to be presented here in full.
\begin{hscode}\SaveRestoreHook
\column{B}{@{}>{\hspre}l<{\hspost}@{}}%
\column{3}{@{}>{\hspre}l<{\hspost}@{}}%
\column{22}{@{}>{\hspre}l<{\hspost}@{}}%
\column{E}{@{}>{\hspre}l<{\hspost}@{}}%
\>[3]{}\mathbf{instance}\;\Conid{Editable}\;{}\<[22]%
\>[22]{}\Conid{Editor}\cdots{}\<[E]%
\\
\>[3]{}\mathbf{instance}\;\Conid{Editable}\;{}\<[22]%
\>[22]{}\Conid{Insertion}\cdots{}\<[E]%
\\
\>[3]{}\mathbf{instance}\;\Conid{Editable}\;{}\<[22]%
\>[22]{}\Conid{Consumption}\cdots{}\<[E]%
\ColumnHook
\end{hscode}\resethooks
This instantiation implies an instance of \ensuremath{\Conid{Patch}\;\Conid{Editor}\;\Conid{Edit}}, which can be
lifted to group words by folding over the \ensuremath{\Conid{Maybe}} monad.
\begin{hscode}\SaveRestoreHook
\column{B}{@{}>{\hspre}l<{\hspost}@{}}%
\column{3}{@{}>{\hspre}l<{\hspost}@{}}%
\column{9}{@{}>{\hspre}c<{\hspost}@{}}%
\column{9E}{@{}l@{}}%
\column{13}{@{}>{\hspre}l<{\hspost}@{}}%
\column{14}{@{}>{\hspre}c<{\hspost}@{}}%
\column{14E}{@{}l@{}}%
\column{18}{@{}>{\hspre}l<{\hspost}@{}}%
\column{E}{@{}>{\hspre}l<{\hspost}@{}}%
\>[3]{}\Varid{semantics}{}\<[14]%
\>[14]{}\mathbin{::}{}\<[14E]%
\>[18]{}\Conid{Word}\;\Conid{Edit}\to \Conid{Editor}{}\<[E]%
\\
\>[3]{}\Varid{semantics}{}\<[14]%
\>[14]{}\mathrel{=}{}\<[14E]%
\>[18]{}\Varid{fromMaybe}\mathbin{\circ}\Varid{foldM}\;\Varid{action}\;\Varid{done}\mathbin{\circ}\Varid{toList}{}\<[E]%
\\[\blanklineskip]%
\>[3]{}\Varid{done}{}\<[9]%
\>[9]{}\mathbin{::}{}\<[9E]%
\>[13]{}\Conid{Insertion}{}\<[E]%
\\
\>[3]{}\Varid{done}{}\<[9]%
\>[9]{}\mathrel{=}{}\<[9E]%
\>[13]{}\Conid{Ins}\;[\mskip1.5mu \mskip1.5mu]\;\Conid{Return}{}\<[E]%
\\[\blanklineskip]%
\>[3]{}\Varid{fromMaybe}{}\<[14]%
\>[14]{}\mathbin{::}{}\<[14E]%
\>[18]{}\Conid{Maybe}\;\Conid{Insertion}\to \Conid{Editor}{}\<[E]%
\\
\>[3]{}\Varid{fromMaybe}{}\<[14]%
\>[14]{}\mathrel{=}{}\<[14E]%
\>[18]{}\Varid{maybe}\;\Conid{Fail}\;\Conid{Try}{}\<[E]%
\ColumnHook
\end{hscode}\resethooks

The adequacy of the semantics can be stated concisely in terms of two
propositions for soundness and full abstraction, respectively.
\begin{hscode}\SaveRestoreHook
\column{B}{@{}>{\hspre}l<{\hspost}@{}}%
\column{3}{@{}>{\hspre}l<{\hspost}@{}}%
\column{E}{@{}>{\hspre}l<{\hspost}@{}}%
\>[3]{}\Varid{semantics\char95 sound}\mathbin{::}\Conid{Meta}\;(\Conid{Word}\;\Conid{Edit}\to \Conid{String}\to \Conid{Bool}){}\<[E]%
\\
\>[3]{}\Varid{semantics\char95 sound}\mathrel{=}\Varid{foreach}\;(\lambda \Varid{x}\to \Varid{patch\char95 eq}\;\Varid{x}\;(\Varid{semantics}\;\Varid{x})){}\<[E]%
\\[\blanklineskip]%
\>[3]{}\Varid{semantics\char95 abstract}\mathbin{::}\Conid{Meta}\;((\Conid{Editor},\Conid{Editor})\to \Conid{Bool}){}\<[E]%
\\
\>[3]{}\Varid{semantics\char95 abstract}\mathrel{=}\Varid{foreach}\;(\Varid{uncurry}\;\Varid{cons\char95 eq}){}\<[E]%
\ColumnHook
\end{hscode}\resethooks
The former uses extensional equivalence under \ensuremath{\Varid{action}} in positive position,
hence the universal quantifiers can be nested, and sampled together at checking
time.
\begin{hscode}\SaveRestoreHook
\column{B}{@{}>{\hspre}l<{\hspost}@{}}%
\column{3}{@{}>{\hspre}l<{\hspost}@{}}%
\column{E}{@{}>{\hspre}l<{\hspost}@{}}%
\>[3]{}\Varid{patch\char95 eq}\mathbin{::}(\Conid{Patch}\;\sigma\;\alpha,\Conid{Patch}\;\sigma\;\beta,\Conid{Eq}\;\sigma)\Rightarrow \alpha\to \beta\to \Conid{Meta}\;(\sigma\to \Conid{Bool}){}\<[E]%
\\
\>[3]{}\Varid{patch\char95 eq}\;\Varid{x}\;\Varid{y}\mathrel{=}\Conid{Meta}\;(\lambda \Varid{s}\to \Varid{action}\;\Varid{s}\;\Varid{x}\equiv \Varid{action}\;\Varid{s}\;\Varid{y}){}\<[E]%
\ColumnHook
\end{hscode}\resethooks
By contrast, the latter conceptually uses extensional equivalence in negative
position: ``if two automata are extensionally equivalent then they are equal''.
This form of quantification cannot be approximated monotonically by sampling.
Hence a constructive solution for the DeMorganized corresponding existential is
required; if two automata are extensionally inequivalent, then a witness state
for which they fail to coincide must be found.
\begin{hscode}\SaveRestoreHook
\column{B}{@{}>{\hspre}l<{\hspost}@{}}%
\column{3}{@{}>{\hspre}l<{\hspost}@{}}%
\column{E}{@{}>{\hspre}l<{\hspost}@{}}%
\>[3]{}\Varid{cons\char95 eq}\mathbin{::}(\Conid{Eq}\;\alpha,\Conid{Witness}\;(\Conid{Diff}\;\alpha)\;\beta)\Rightarrow \alpha\to \alpha\to \Conid{Meta}\;\Conid{Bool}{}\<[E]%
\\
\>[3]{}\Varid{cons\char95 eq}\;\Varid{t}\;\Varid{u}\mathrel{=}\Conid{Meta}\;(\Varid{t}\equiv \Varid{u}\mathrel{\vee}\Varid{existsSome}\;(\Varid{t}\mathrel{{:}{\not\equiv}}\Varid{u})){}\<[E]%
\ColumnHook
\end{hscode}\resethooks
The operator \ensuremath{\mathrel{{:}{\not\equiv}}} of constructive logic is conveniently defined as the
constructor of a new datatype \ensuremath{\Conid{Diff}}, since it requires an ad-hoc instance of
\ensuremath{\Conid{Witness}} to hold each construction algorithm.
\begin{hscode}\SaveRestoreHook
\column{B}{@{}>{\hspre}l<{\hspost}@{}}%
\column{3}{@{}>{\hspre}l<{\hspost}@{}}%
\column{E}{@{}>{\hspre}l<{\hspost}@{}}%
\>[3]{}\mathbf{data}\;\Conid{Diff}\;\alpha\mathrel{=}(\mathrel{{:}{\not\equiv}})\;\alpha\;\alpha{}\<[E]%
\ColumnHook
\end{hscode}\resethooks
It turns out that a straightforward algorithm requires three auxiliary
constructive predicates, which bear witness that an automaton accepts some
input, that an automaton rejects some input, and that one automaton accepts an
input whereas another one rejects it, respectively.
\begin{hscode}\SaveRestoreHook
\column{B}{@{}>{\hspre}l<{\hspost}@{}}%
\column{3}{@{}>{\hspre}l<{\hspost}@{}}%
\column{24}{@{}>{\hspre}c<{\hspost}@{}}%
\column{24E}{@{}l@{}}%
\column{27}{@{}>{\hspre}l<{\hspost}@{}}%
\column{E}{@{}>{\hspre}l<{\hspost}@{}}%
\>[3]{}\mathbf{data}\;\Conid{Def}\;\alpha{}\<[24]%
\>[24]{}\mathrel{=}{}\<[24E]%
\>[27]{}\Conid{Def}\;\alpha{}\<[E]%
\\[\blanklineskip]%
\>[3]{}\mathbf{data}\;\Conid{Undef}\;\alpha{}\<[24]%
\>[24]{}\mathrel{=}{}\<[24E]%
\>[27]{}\Conid{Undef}\;\alpha{}\<[E]%
\\[\blanklineskip]%
\>[3]{}\mathbf{data}\;\Conid{DefUndef}\;\alpha{}\<[24]%
\>[24]{}\mathrel{=}{}\<[24E]%
\>[27]{}(\mathrel{{:}{\geq}})\;\alpha\;\alpha{}\<[E]%
\ColumnHook
\end{hscode}\resethooks
The implementations are too complex to be discussed here in detail.
\begin{hscode}\SaveRestoreHook
\column{B}{@{}>{\hspre}l<{\hspost}@{}}%
\column{3}{@{}>{\hspre}l<{\hspost}@{}}%
\column{31}{@{}>{\hspre}l<{\hspost}@{}}%
\column{E}{@{}>{\hspre}l<{\hspost}@{}}%
\>[3]{}\mathbf{instance}\;\Conid{Witness}\;(\Conid{Diff}\;{}\<[31]%
\>[31]{}\Conid{Editor})\;\Conid{String}\cdots{}\<[E]%
\\
\>[3]{}\mathbf{instance}\;\Conid{Witness}\;(\Conid{Def}\;{}\<[31]%
\>[31]{}\Conid{Editor})\;\Conid{String}\cdots{}\<[E]%
\\
\>[3]{}\mathbf{instance}\;\Conid{Witness}\;(\Conid{Undef}\;{}\<[31]%
\>[31]{}\Conid{Editor})\;\Conid{String}\cdots{}\<[E]%
\\
\>[3]{}\mathbf{instance}\;\Conid{Witness}\;(\Conid{DefUndef}\;{}\<[31]%
\>[31]{}\Conid{Editor})\;\Conid{String}\cdots{}\<[E]%
\ColumnHook
\end{hscode}\resethooks
However, the intended semantics can be specified precisely, and checked, in a
self-application of the meta-logical language.
\begin{hscode}\SaveRestoreHook
\column{B}{@{}>{\hspre}l<{\hspost}@{}}%
\column{3}{@{}>{\hspre}l<{\hspost}@{}}%
\column{5}{@{}>{\hspre}l<{\hspost}@{}}%
\column{53}{@{}>{\hspre}l<{\hspost}@{}}%
\column{E}{@{}>{\hspre}l<{\hspost}@{}}%
\>[3]{}\Varid{def\char95 sound\char95 complete}\mathbin{::}\Conid{Meta}\;(\Conid{Editor}\to \Conid{Bool}){}\<[E]%
\\
\>[3]{}\Varid{def\char95 sound\char95 complete}\mathrel{=}\Conid{Meta}{}\<[E]%
\\
\>[3]{}\hsindent{2}{}\<[5]%
\>[5]{}(\lambda \Varid{x}\to \Varid{x}\equiv \Conid{Fail}\mathrel{\vee}\Varid{exists}\;(\Conid{Def}\;\Varid{x})\;(\lambda \Varid{s}\to \Varid{isJust}\;(\Varid{action}\;\Varid{s}\;\Varid{x}))){}\<[E]%
\\[\blanklineskip]%
\>[3]{}\Varid{undef\char95 sound\char95 complete}\mathbin{::}\Conid{Meta}\;(\Conid{Editor}\to \Conid{Bool}){}\<[E]%
\\
\>[3]{}\Varid{undef\char95 sound\char95 complete}\mathrel{=}\Conid{Meta}{}\<[E]%
\\
\>[3]{}\hsindent{2}{}\<[5]%
\>[5]{}(\lambda \Varid{x}\to \Varid{isTotal}\;\Varid{x}\mathrel{\vee}\Varid{exists}\;(\Conid{Undef}\;\Varid{x})\;(\lambda \Varid{s}\to \neg \;(\Varid{isJust}\;(\Varid{action}\;\Varid{s}\;\Varid{x})))){}\<[E]%
\\[\blanklineskip]%
\>[3]{}\Varid{def\char95 undef\char95 sound}\mathbin{::}\Conid{Meta}\;((\Conid{Editor},\Conid{Editor})\to \Conid{Bool}){}\<[E]%
\\
\>[3]{}\Varid{def\char95 undef\char95 sound}\mathrel{=}\Conid{Meta}{}\<[E]%
\\
\>[3]{}\hsindent{2}{}\<[5]%
\>[5]{}(\lambda (\Varid{x},\Varid{y})\to \Varid{existsOrVacuous}\;(\Varid{x}\mathrel{{:}{\geq}}\Varid{y})\;(\lambda \Varid{s}\to {}\<[53]%
\>[53]{}\Varid{isJust}\;(\Varid{action}\;\Varid{s}\;\Varid{x})\mathrel{\wedge}{}\<[E]%
\\
\>[53]{}\neg \;(\Varid{isJust}\;(\Varid{action}\;\Varid{s}\;\Varid{y})))){}\<[E]%
\\[\blanklineskip]%
\>[3]{}\Varid{diff\char95 sound\char95 complete}\mathbin{::}\Conid{Meta}\;((\Conid{Editor},\Conid{Editor})\to \Conid{Bool}){}\<[E]%
\\
\>[3]{}\Varid{diff\char95 sound\char95 complete}\mathrel{=}\Conid{Meta}{}\<[E]%
\\
\>[3]{}\hsindent{2}{}\<[5]%
\>[5]{}(\lambda (\Varid{x},\Varid{y})\to \Varid{x}\equiv \Varid{y}\mathrel{\vee}\Varid{exists}\;(\Varid{x}\mathrel{{:}{\not\equiv}}\Varid{y})\;(\lambda \Varid{s}\to \Varid{action}\;\Varid{s}\;\Varid{x}\not\equiv \Varid{action}\;\Varid{s}\;\Varid{y})){}\<[E]%
\ColumnHook
\end{hscode}\resethooks
Note that the target \ensuremath{\Varid{semantics\char95 abstract}} is a consequence of
\ensuremath{\Varid{diff\char95 sound\char95 complete}} a fortiori already, but there is no obvious way to exploit
that logical relationship in a checking framework.

Now we can operationalize the word problem by comparing automata,
\begin{hscode}\SaveRestoreHook
\column{B}{@{}>{\hspre}l<{\hspost}@{}}%
\column{3}{@{}>{\hspre}l<{\hspost}@{}}%
\column{E}{@{}>{\hspre}l<{\hspost}@{}}%
\>[3]{}(\cong)\mathbin{::}\Conid{Word}\;\Conid{Edit}\to \Conid{Word}\;\Conid{Edit}\to \Conid{Bool}{}\<[E]%
\\
\>[3]{}\Varid{x}\cong\Varid{y}\mathrel{=}\Varid{semantics}\;\Varid{x}\equiv \Varid{semantics}\;\Varid{y}{}\<[E]%
\ColumnHook
\end{hscode}\resethooks
and conclude for instance that \ensuremath{\Varid{fromList}\;[\mskip1.5mu \Conid{Edit}\;\Conid{Insert}\;\mathrm{2}\;\text{\ttfamily 'a'},\Conid{Edit}\;\Conid{Delete}\;\mathrm{3}\;\text{\ttfamily 'b'}\mskip1.5mu]}
\ensuremath{\cong} \ensuremath{\Varid{fromList}\;[\mskip1.5mu \Conid{Edit}\;\Conid{Delete}\;\mathrm{2}\;\text{\ttfamily 'b'},\Conid{Edit}\;\Conid{Insert}\;\mathrm{2}\;\text{\ttfamily 'a'}\mskip1.5mu]}.

\subsection{Strategical Remarks}

The above examples let us have a glimpse at the power of recursive tactics
available in an embedded higher-order logical language: Equational reasoning
about the data to be modeled is reduced to the word problem of a group
presentation, an extensional property quantified over all possible inputs that
can only be checked heuristically and positively.  For broader checkability,
this problem is factored through a normalizable automaton representation.  The
soundness and full abstraction of this semantics, and thus the equivalence of
its equational reasoning, are checkable properties that contain existential
quantification, for which constructive witnesses are given.  The correctness and
completeness of these constructions are universally quantified properties again,
which can be checked heuristically.

Note that this does not mean we are going in circles; the correctness of the
semantics needs only to be established once, and can be used as a shortcut for
deciding equations of the original model henceforth.  Yet the same language,
tools and workflow are used for all phases.

\section{Conclusion}

We have proposed three advanced features of meta-logical language for offline
checking of functional programs, namely meta-level marking, nominal axiomatics
and constructive existentials.  We have shown their implementation in the Haskell
checking framework PureCheck, and demonstrated their use and interaction by means of a
nontrivial executable modeling problem.
\REV{Other aspects of PureCheck, such as ensuring the efficiency of deterministic sampling, are both work in progress and out of scope here, and shall be discussed in a forthcoming companion paper.}

Dialectically, the stylistic ideal that underlies our experiments is contrary to
the one employed in the construction of this paper\REV{:}  The checking paradigm
expresses reasoning about the program in (a marked level of) the code, as
opposed to the prose \REV{embellishment} of the literate paradigm.  We are hopeful that a
thorough synthesis of the two can be demonstrated as synergetic and useful in
the future.

Expressive offline checking language is an important step towards the
reification of the algebraic concepts that pervade functional program design;
consider the ubiquitous informal equational theory associated with Haskell
type classes.

Marking the \ensuremath{\Conid{Meta}} level explicitly has not only the demonstrated advantages for
the human reader, but may also serve as an anchor for meta-programming
procedures, such as automatic test suite extraction \REV{without magic names}.

The concept of constructive existentials is an explicitly controlled counterpart
to implicit search strategies provided by the logical programming paradigm.
\REV{Unlike SmartCheck, where constructive existentials are dismissed for often being hard to find in practice, we contend that in a (self-)educational context such as executable modeling, the understanding gained by implementing the construction witnesses for existential meta-logical properties of interest is rewarding rather than onerous.} Furthermore
foresee interesting potential in the transfer of our ideas to a
functional--logic language such as Curry\,\cite{curry} \REV{with built-in encapsulated search capabilities}, but leave the
exploration for future work.


\end{document}